\documentclass{elsart}
\usepackage{graphicx}
\usepackage{ifthen}             

\newcommand{\Lnoise}[1]{{\mathcal L}_{#1}}   
\newcommand{\Lmat}[1]{{{\bf L}_{#1}}}	   
\newcommand{\Bmat}[1]{{{\bf B}_{#1}}}	   
\newcommand{\Prpgtr}[1]{\Delta_{#1}}
\newcommand{\InvPrpgtr}[1]{\Delta^{-1}_{#1}}

\newcommand{\field}{\phi}
\newcommand{\Df}[1]{f^{'}_{#1}}

\newcommand{\mod}[1]{#1}

\newcommand{\PC}[1]{$\footnotemark\footnotetext{PC: #1}$}

\newcommand{\rf}     [1] {~\cite{#1}}
\newcommand{\refref} [1] {ref.~\cite{#1}}

\newcommand{\refeq}  [1] {(\ref{#1})}

\newcommand{\reffig} [1] {figure~\ref{#1}}

\newcommand{\refsect}[1] {sect.~\ref{#1}}

\newcommand{\beq}{\begin{equation}}
\newcommand{\continue}{\nonumber \\ }
\newcommand{\nnu}{\nonumber}
\newcommand{\eeq}{\end{equation}}
\newcommand{\ee}[1] {\label{#1} \end{equation}}
\newcommand{\bea}{\begin{eqnarray}}

\newcommand{\eea}{\end{eqnarray}}
\newcommand{\barr}{\begin{array}}
\newcommand{\earr}{\end{array}}

\newcommand{\btrack}[1]{\raisebox{-2.0ex}[3.5ex][2.5ex]
        {\includegraphics[height=5ex]{#1.eps}\negthinspace} }
\newcommand{\btrackA}[1]{\raisebox{-3.0ex}[4.5ex][3.5ex]
         {\includegraphics[height=7ex]{#1.eps}\negthinspace} }

\newlength{\Fsize}   
\newcommand{\FIG}[4]{\begin{figure}
                      {#1}
                      \caption[#2]{#3}
                      \label{#4} \end{figure} }

\newcommand{\evOper}{evolution oper\-ator}
\newcommand{\EvOper}{Evolution oper\-ator}
\newcommand{\FPoper}{Perron-Frobenius oper\-ator} 

\newcommand{\dzeta}{dyn\-am\-ic\-al zeta func\-tion}

\newcommand{\fd}{spec\-tral det\-er\-min\-ant}

\newcommand{\obser}{a}		
\newcommand{\Obser}{A}		

\newcommand{\pde}{\partial}

\renewcommand{\det}{\mbox{\rm det}}

\newcommand{\tr}{{\rm tr}\, }

\newcommand{\prpgtr}[1]{\delta\negthinspace\left( {#1} \right)}

\newcommand{\Lop}{{\mathcal L}}	   
\newcommand{\oneMinJ}[1]{\left|\det\left({\bf 1}-{\bf J}_p^{#1}\right)\right|}
\newcommand{\ExpaEig}{\Lambda}	   
\newcommand{\eigenvL}{{\nu}}       
\newcommand{\inFix}[1]{{\in \mbox{\footnotesize Fix}f^{#1}}}

\newcommand\period[1]{{T_{#1}}}			
\newcommand{\cl}[1]{{n_{#1}}}	

\begin{document}
\runauthor{Cvitanovi\c'}
\begin{frontmatter}
\title{Chaotic Field Theory: a Sketch}
\author{Predrag Cvitanovi\'c} 

\address{
Department of Physics \&\ Astronomy, Northwestern University \\
2145 Sheridan Road, Evanston, Illinois 60208
}


\begin{abstract}
Spatio-temporally chaotic dynamics of a
classical field can be described  
by means of an infinite hierarchy of its unstable spatio-temporally
periodic solutions. 
The periodic orbit theory yields  the
global averages characterizing the chaotic dynamics, as well
as the starting semiclassical approximation to the quantum theory.

New methods for computing corrections to the semiclassical
approximation are developed; in particular, a nonlinear field
transformation yields 
the perturbative corrections in a form more compact than
the Feynman diagram expansions.
\end{abstract}

\end{frontmatter}


Formulated in 1946-49 and tested through 1970's, quantum electrodynamics takes
free electrons and photons as its point of departure, 
with nonlinear effects taken
in account perturbatively in terms of Feynman diagrams, as corrections
of order $({\alpha / \pi})^n = (0.002322819\dots)^n$.
QED is a wildly successful theory, with Kinoshita's\rf{KinoHughes} 
calculation of the electron magnetic moment
\[
{1 \over 2} (g-2) = \sum_n \left({\alpha \over \pi}\right)^n
\left\{ \cdots \,+\, \btrackA{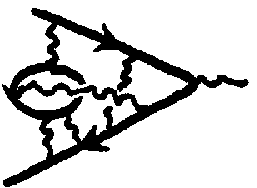}  \,+\,\cdots \right\}
\]
agreeing with Dehmelt's experiments\rf{Dehmelt87} to 
12 significant digits.

Quantum chromodynamics perturbative calculations
seemed the natural next step,
the only new feature being the gluon-gluon interactions.
However, in this case the Feynman--diagrammatic expansions 
for observables such as the meson and hadron masses
\[
\mbox{(observable)} = \sum_n \left(\alpha_{QCD}\right)^n
\left\{ \cdots \,+\, \btrackA{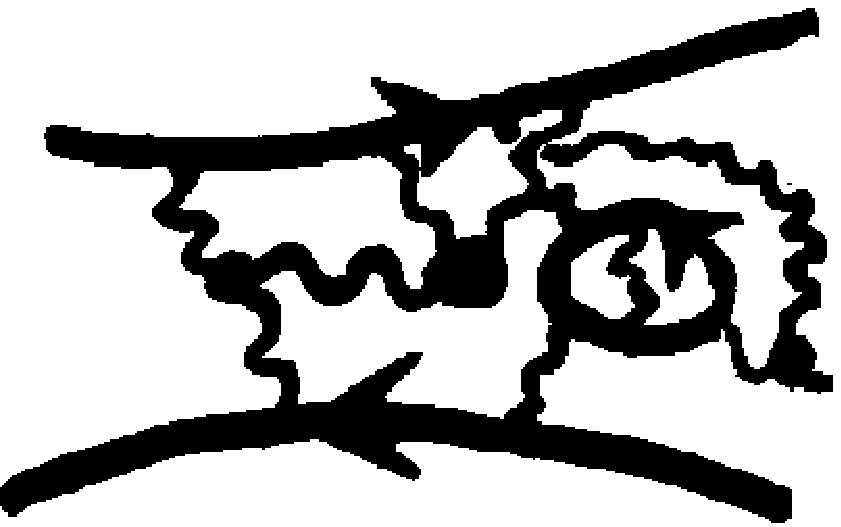} \,+\,\cdots \right\}
\]
failed us utterly, perhaps because
the expansion parameter is of order 1. I say perhaps,
because more likely the error in
this case  is thinking in terms of quarks and gluons
in the first place. Strongly nonlinear field theories
require radically different approaches, 
and in 1970's, with a deeper appreciation of the connections between 
field theory and statistical mechanics, their re-examination led
to path integral formulations such as the lattice QCD\rf{KWilson}.
In lattice theories quantum fluctuations explore the full gauge group
manifold, and classical dynamics of Yang-Mills fields plays no role.

We propose to re-examine here the path integral formulation
and the role that the classical solutions play
in quantization of strongly nonlinear fields.
In the path integral formulation of a field theory the dominant 
contributions come from saddlepoints, the classical solutions
of equations of motion. Usually one imagines one
dominant saddle point, the ``vacuum'':
\vspace{2ex}
\\
{\centerline{
	${
	\hspace{-4ex}
	\includegraphics[width=0.30\textwidth]{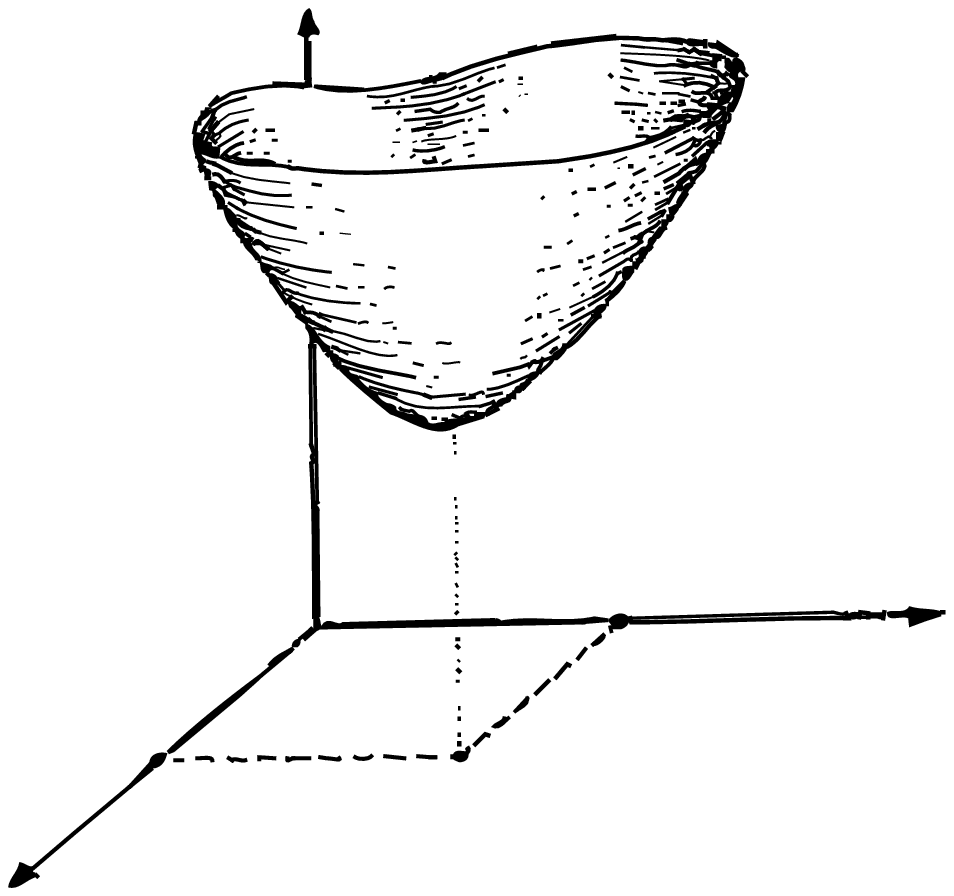}
	\hspace{-4ex}
	\atop
	\mbox{one dominant extremum}
	}$
~~~~
	${
	\hspace{-2ex}
		 {
	   \includegraphics[width=0.28\textwidth]{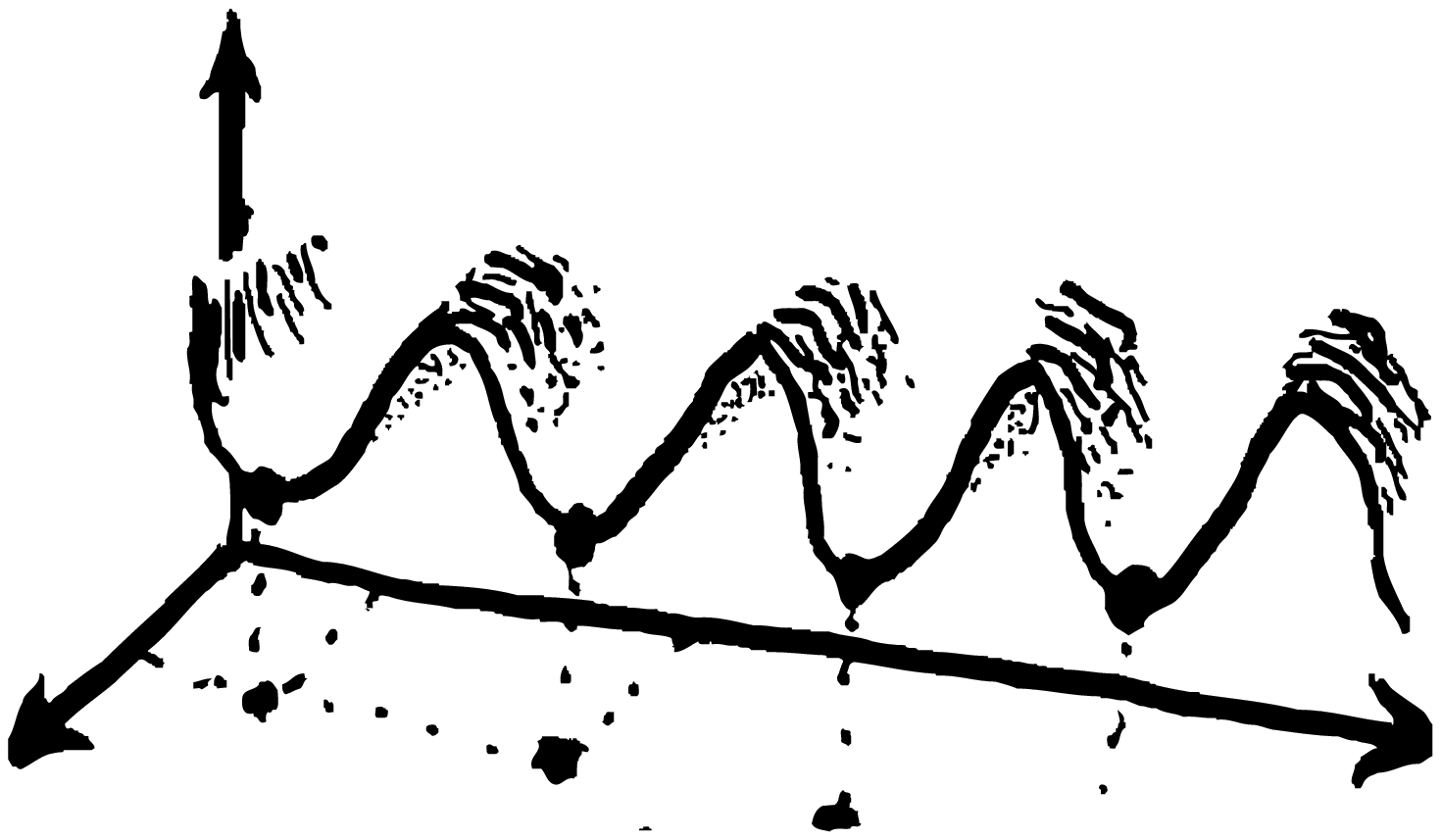}
		 }
	\hspace{-2ex}
	\atop
	\mbox{an infinity of instanton saddles}
	}$
          }}
\vspace{1ex}
\\
The Feynman diagrams of QED and QCD are
nothing more than a scheme to compute the correction terms to 
this starting 
semiclassical, Gaussian saddlepoint approximation.
But there might be other saddles.
That field theories might have
a rich repertoire of classical solutions became apparent
with the discovery of instantons\rf{Polyakov}, analytic
solutions of the classical $SU(2)$ Yang-Mills equations of
motion, and the realization that the associated instanton vacua
receive contributions from countable $\infty$'s of
saddles. What is not clear is whether these are the important
classical saddles. Could it be that the strongly nonlinear theories 
are dominated by altogether different
classical solutions?

The search for the classical
solutions of nonlinear field theories such as the Yang-Mills and
gravity has so far been neither very successful nor very systematic.
In modern field theories the main emphasis has been
on symmetries
as guiding principles in writing down the actions. 
But writing down a differential equation is only the start of the story;
even for systems as simple as
3 coupled ordinary differential equations one in general 
has no clue what the nature of the long time solutions might be.

These are hard problems, and 
in explorations of modern field theories the dynamics tends to
be is neglected, and
understandably so, because the wealth of the classical solutions
of nonlinear systems can be truly bewildering.
If the classical behavior of these theories is anything like
that of 
the field theories that describe the classical world ---
the hydrodynamics, the 
magneto-hydrodynamics, the
Ginzburg-Landau
system
---
there should be very many solutions, with very few of the important
ones analytical in form; the strongly nonlinear classical field theories
are turbulent, after all.
Furthermore, there is not a dimmest hope that such solutions
are either beautiful or analytic,
and there is not much enthusiasm for grinding out
numerical solutions
as long as one lacks ideas as what to do with them.

By late 1970's it was generally
understood that even the simplest nonlinear systems
exhibit chaos. Chaos is the norm also for generic
Hamiltonian flows, and for path integrals that implies that 
instead of a few, or countably few saddles, 
classical solutions populate 
fractal sets of saddles. 
\vspace{1ex}
\\
\centerline{
 ${
	\hspace{-4ex}
	\includegraphics[width=0.40\textwidth]{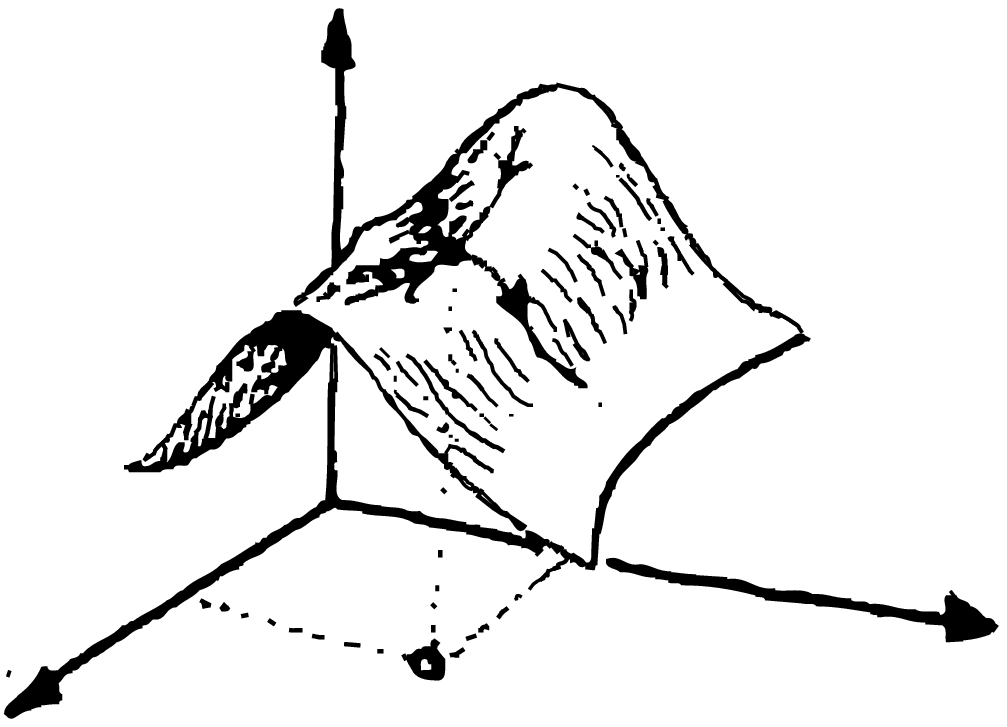}
	\hspace{-4ex}
        \atop
        \mbox{a local unstable extremum}
        }$
~~~~~~~~
 ${
	\includegraphics[width=0.35\textwidth]{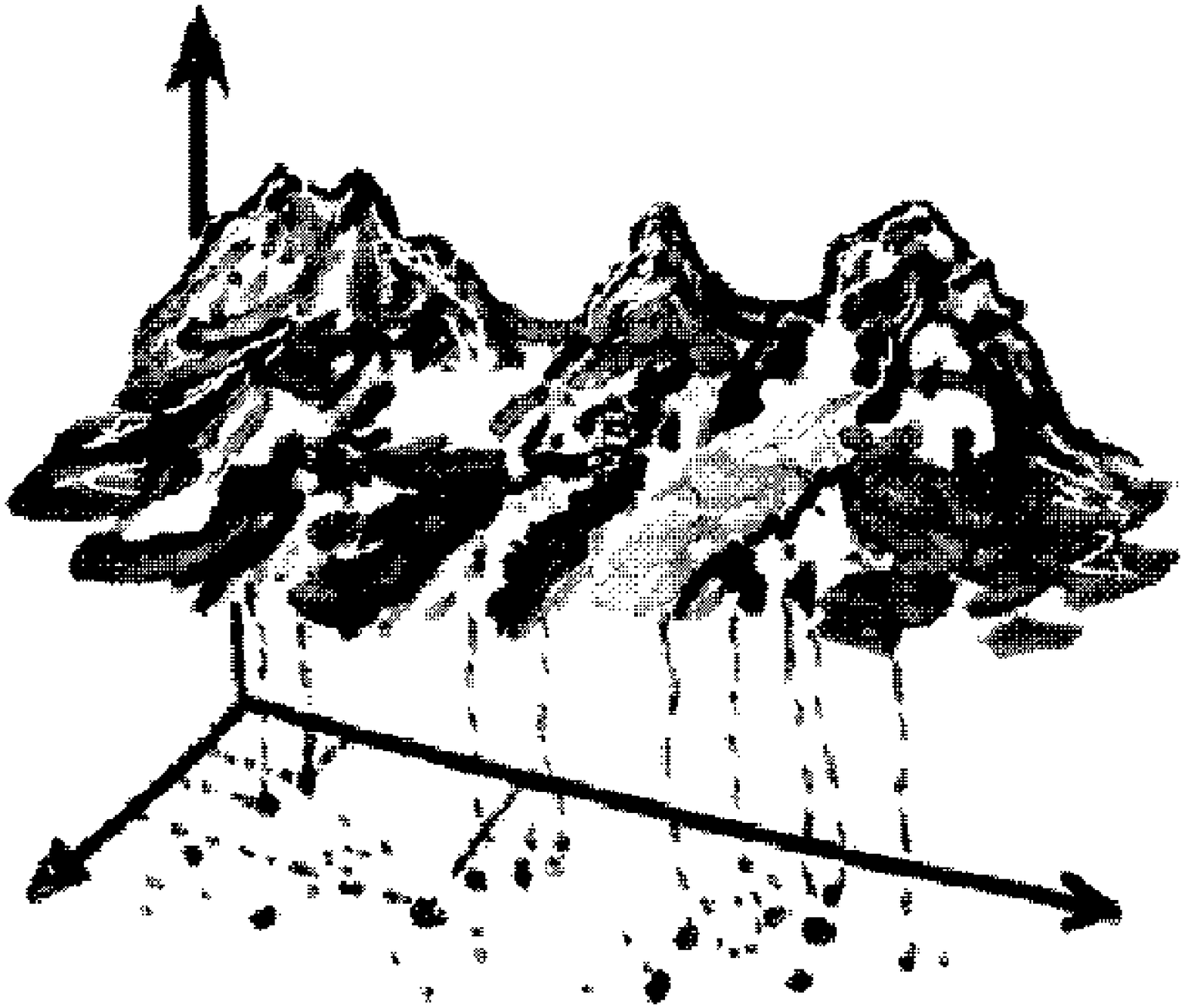}
        \atop
        \mbox{a fractal set of saddles}
        }$
          }
\vspace{1ex}
\\
For the path-integral formulation of quantum mechanics
such solutions 
were discovered and accounted for by Gutzwiller\rf{gutbook}
in late 1960's.
In this framework the spectrum of the theory is computed from
a set of its unstable classical periodic solutions.
The new aspect is that the individual saddles for classically 
chaotic systems are nothing like the harmonic oscillator degrees
of freedom, the quarks and gluons of QCD --- they are all
unstable and highly nontrivial, accessible  only by numerical techniques.

So, if one is to develop a semiclassical field theory of systems that
are classically chaotic or ``turbulent'',
the problem one faces is twofold
\begin{enumerate}
\item
determine, classify, and order by relative importance
the classical solutions of nonlinear field theories.

\item
develop methods for calculating perturbative corrections
to the corresponding classical saddles.

\end{enumerate}

Our purpose here is to give an overview over the status of
this program --- for details the reader is referred to
the literature cited.

The first task, a systematic exploration of solutions
of field theory has so far been implemented only  for one of the very
simplest field theories, the 1-dimensional Kuramoto-Sivashinsky system.
We sketch below how its spatio-temporally chaotic dynamics can be
described in terms of spatio-temporally recurrent
unstable patterns.

For the second task, the theory of perturbative corrections, we shall turn to
an even simpler system;  a weakly stochastic mapping
in 1-dimension.
The new aspect of the theory is that now
the corrections have to be computed saddle by saddle. 
In \refsect{s-PerCorr} to \refsect{scfpo} we discuss
three distinct methods for their evaluation.

\section{Unstable recurrent patterns in classical field theories}

Field theories such as
4-dimensional QCD or gravity have many dimensions, symmetries,
tensorial indices. They are
far too complicated for exploratory forays into 
this forbidding terrain. 
We start instead by taking a simple spatio-temporally chaotic
nonlinear system of physical interest, and investigate the
nature of its solutions.


One of the simplest and extensively studied
spatially extended dynamical systems is 
the Kuramoto-Sivashinsky system\rf{KurSiv}
\beq
u_t=(u^2)_x-u_{xx}-\nu u_{xxxx} 
\ee{ks}
which arises as an amplitude equation for interfacial instabilities
in a variety of contexts.
The ``flame front'' $u(x,t)$ has compact support, with
$x \in [0,2\pi]$ a periodic space coordinate.
The $u^2$ term makes this a nonlinear system,
$t$
is the time, 
and $\nu$ is a fourth-order ``viscosity'' damping parameter
that irons out any sharp features.
Numerical simulations demonstrate that as
the viscosity decreases (or the size of the system increases),
the ``flame front'' becomes increasingly unstable and turbulent.
The task of the theory is to describe this spatio-temporal
turbulence and yield quantitative predictions for its measurable
consequences.

Armed with a computer and a great deal of skill,
one can obtain a numerical solution to a nonlinear
PDE. The real question is;
once a solution is found, what is to be done with it?
The periodic orbit theory is an answer to this question.

Dynamics drives a given spatially extended system through a
repertoire of unstable patterns; as we watch  
a ``turbulent'' system evolve, 
every so often we catch a glimpse of a familiar pattern: 
\vspace{4ex}
\\
\centerline{
	\raisebox{-4.0ex}[5.5ex][4.5ex]
		 {
\includegraphics[height=12ex]{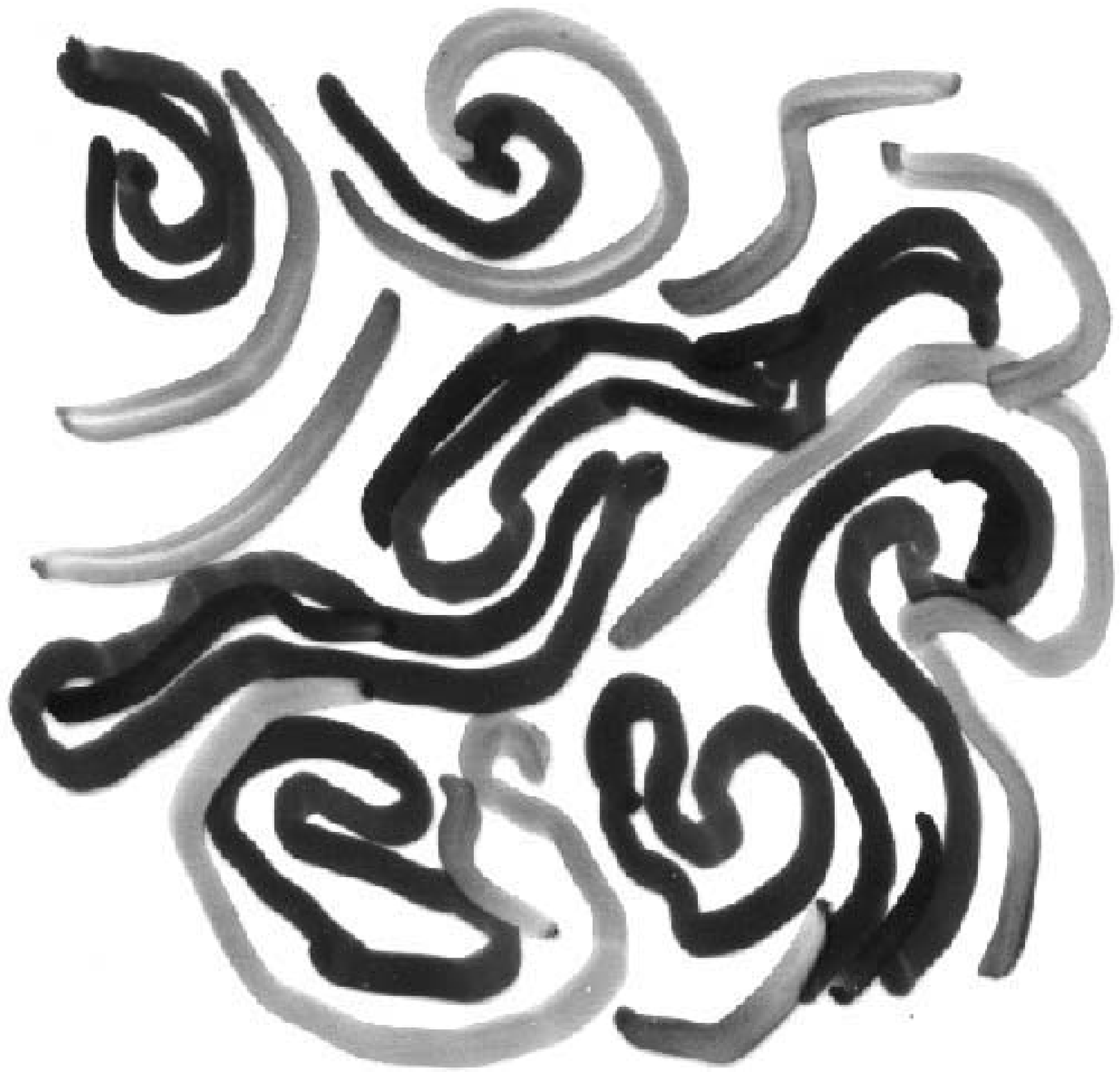} 
		 }
~~~
$\Longrightarrow$
~~
{other swirls}
~~
$\Longrightarrow$
~~~
	\raisebox{-4.0ex}[5.5ex][4.5ex]
		 {
\includegraphics[height=12ex]{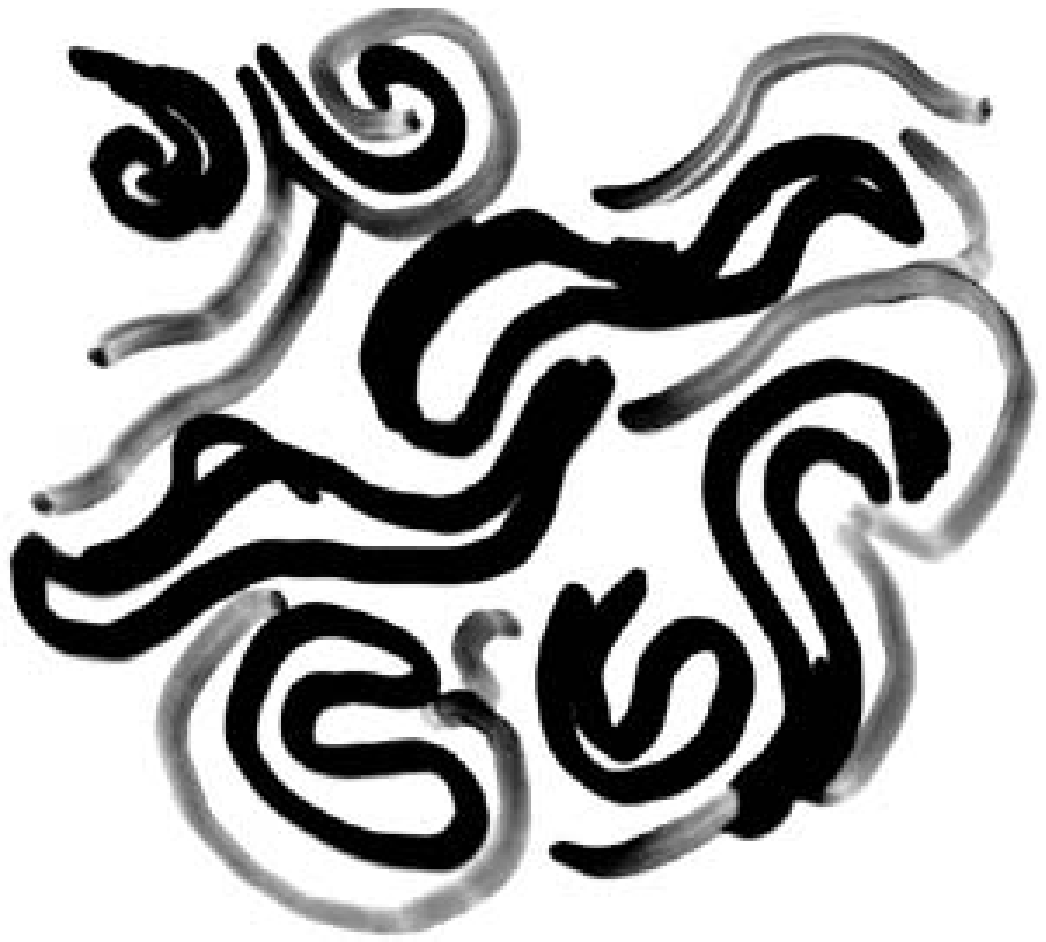}
		 }
          }
\vspace{1ex}
\\
For any finite  spatial resolution,
the system follows approximately for a finite time 
a pattern belonging to a finite 
alphabet of admissible patterns, and the long term dynamics can be thought
of as a walk through the space of such patterns,
just as chaotic dynamics with a  low dimensional
attractor can be thought of as a succession of nearly periodic (but
unstable) motions.
The periodic orbit provides the
machinery that converts this intuitive picture into
precise calculation scheme that extracts asymptotic time predictions
from the short time dynamics.
For extended systems the theory
gives a description of the asymptotics of
partial differential equations 
in terms of recurrent spatio-temporal patterns.

Putkaradze has proposed that
the Kuramoto-Sivashinsky system \refeq{ks}
be used as a laboratory for exploring such ideas.
We now summarize the results obtained so far in this direction
by 
Christiansen et al.\rf{CCP96} and
Zoldi and Greenside\rf{ZG96}.

The solution $u(x,t)=u(x+2\pi,t)$ is periodic on 
the  $x \in [0,2\pi]$ interval, so one (but by no means only)
way to solve such equations is  
to expand $u(x,t)$ in a discrete spatial Fourier series 
\begin{equation}
 u(x,t)= i \sum_{k=-\infty}^{+ \infty} a_k(t) e^{i k x}
\, . 
\label{fseries}
\end{equation}
Restrict the consideration
to the subspace of odd solutions $u(x,t)=-u(-x,t)$ for which 
$a_k$ are real. 
Substitution of (\ref{fseries}) into (\ref{ks}) yields 
the infinite ladder of evolution equations for the Fourier coefficients $a_k$: 
\begin{equation}
\dot{a}_k=(k^2-\nu k^4)a_k - k \sum_{m=-\infty}^{\infty} a_m a_{k-m} 
\,.
\label{expan}
\end{equation}
$u(x,t)=0$ is  a fixed point of (\ref{ks}), with
the $k^2 \nu<1$
long wavelength modes of this fixed point 
linearly unstable, and the 
short wavelength modes stable.  
For $\nu > 1$,  $u(x,t)=0$ is the globally attractive stable fixed point;
starting with $\nu =1$ the solutions go through a rich sequence of
bifurcations,
and myriad unstable periodic solutions whose number
grows exponentially with time.

The essential limitation on 
the numerical studies undertaken so far have been 
computational constraints: in
truncation of high modes in the expansion (\ref{expan}), 
sufficiently many have to be retained to ensure 
the dynamics is accurately represented.
Christiansen et al.\rf{CCP96} have
examined the dynamics for values of the damping parameter
close to the onset of chaos, while
Zoldi and Greenside\rf{ZG96}
have explored somewhat more turbulent values of $\nu$.
With improvement of numerical codes
considerably more turbulent regimes should become accessible.

One pleasant surprise is that
even though one is dealing with (infinite
dimensional) PDEs, for these
strong dissipation values of parameters the 
spatio-temporal chaos is sufficiently weak that the
flow can be visualised as an approximately 1-dimensional Poincar\'e return map
$s \rightarrow f(s)$ 
from the unstable manifold of the shortest periodic point
onto its neighborhood, see \reffig{unfolded}(a). 
This representation makes it possible to
systematically determine all nearby periodic solutions up to a given
maximal period.
%
\FIG{
(a)
\hspace{-4ex}
\includegraphics[width=0.48\textwidth]{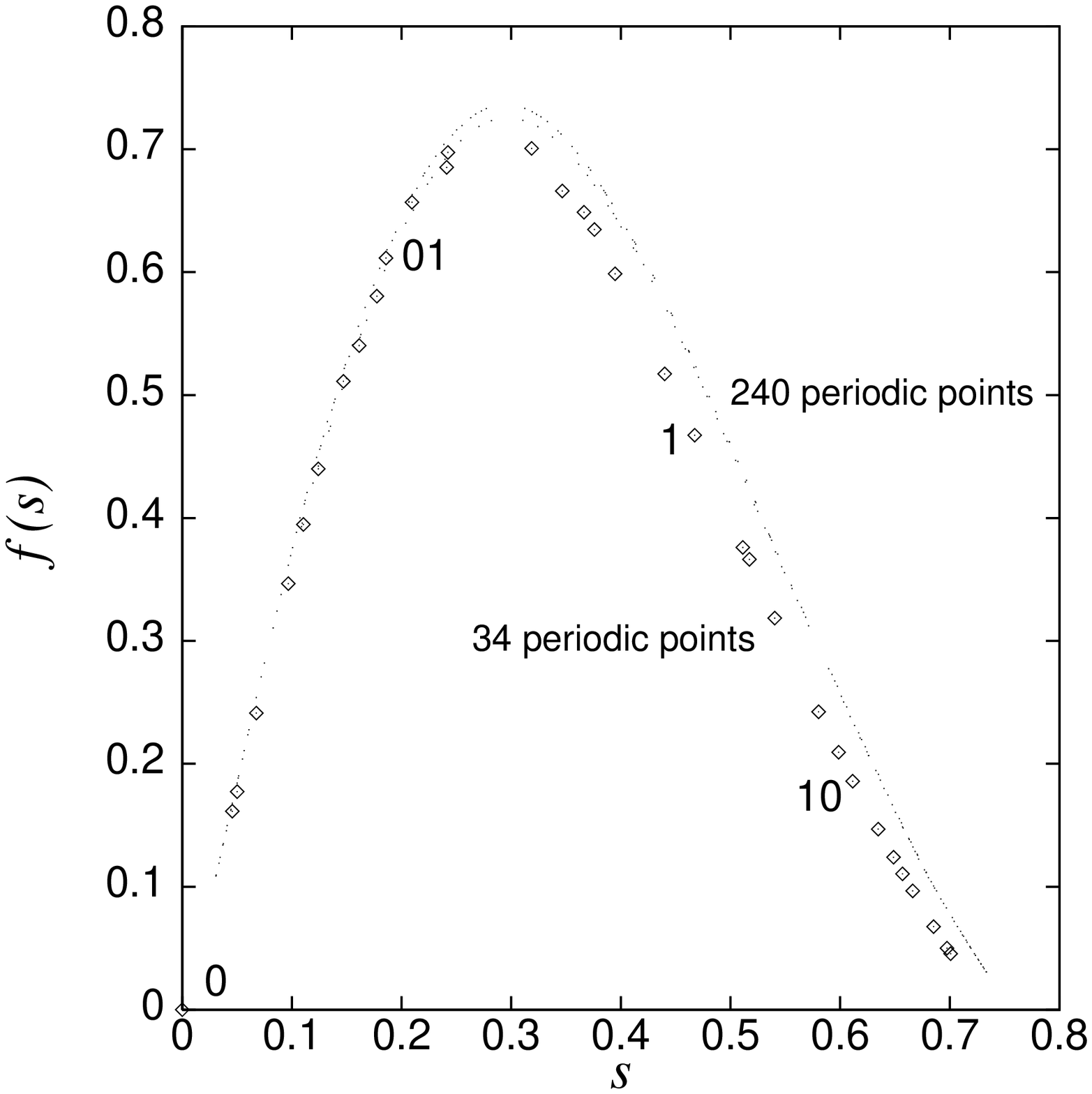}
\hspace{-4ex}
~~~~~~~~
(b)
\hspace{-4ex}
\includegraphics[width=0.48\textwidth]{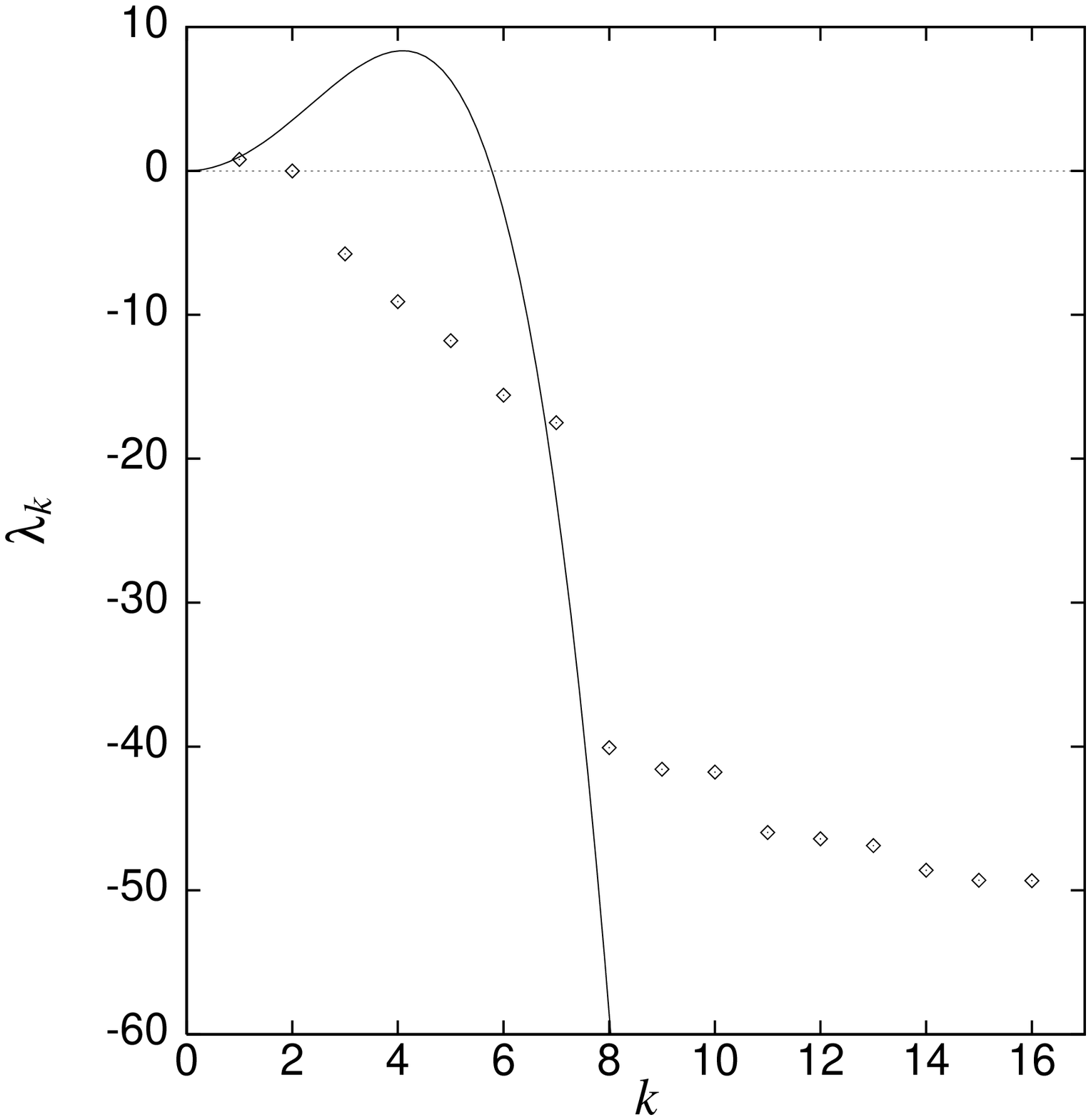}
\hspace{-4ex}
}{}{
(a) The return map $s_{n+1} = f(s_n)$ constructed
from periodic solutions of the Kuramoto-Sivashinsky 
equations \refeq{ks}, $\nu=0.029910$,
with $s$ the distance
measured along the unstable manifold of the fixed
point $\overline{1}$. 
Periodic points $\overline{0}$ 
and $\overline{01}$ are also indicated.
(b) Lyapunov exponents $\lambda_k$ versus $k$ for the 
periodic orbit $\overline{1}$ compared with  the stability eigenvalues 
of the $u(x,t)=0$ stationary solution $k^2- \nu k^4$.  
$\lambda_k$ for $k \geq 8$ lie below the numerical accuracy of integration 
and are not meaningful. 
From \refref{CCP96}.
}{unfolded}
%

So far some 1,000 prime cycles have been determined numerically
for various values of viscosity. 
The rapid contraction in the nonleading eigendirections
is illustrated in \reffig{unfolded}(b) 
by the plot of the first 16 eigenvalues of the 
$\overline{1}$-cycle.
As the length of the orbit increases, the magnitude of
contracting eigenvalues falls off very quickly.
In \reffig{orbit0fig} we plot 
$u_0(x,t)$ corresponding to the $\overline{0}$-cycle.
The difference between this solution and the other shortest period
solution is of the order of 50\% of a typical variation in
the amplitude of $u(x,t)$, so the chaotic dynamics is 
already exploring a sizable swath in the space of possible
patterns even so close to the onset of spatio-temporal chaos.
Other solutions, plotted in the configuration space, exhibit the same
overall gross structure. 
Together they form the
repertoire of the recurrent spatio-temporal patterns that is being
explored by the turbulent dynamics.

\FIG{
(a)
\hspace{-4ex}
\includegraphics[width=0.50\textwidth]{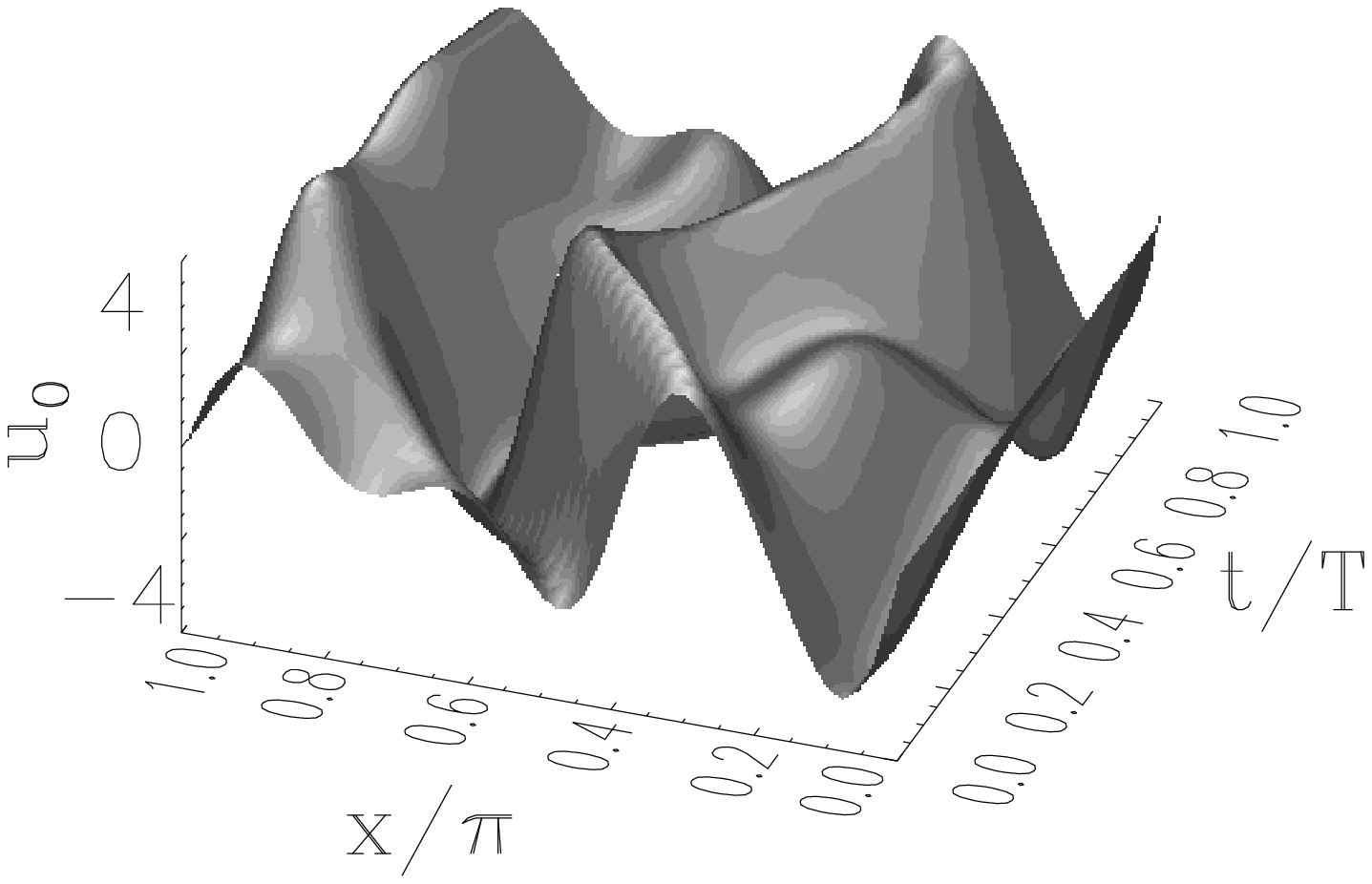}
\hspace{-4ex}
(b)
\hspace{-4ex}
\includegraphics[width=0.50\textwidth]{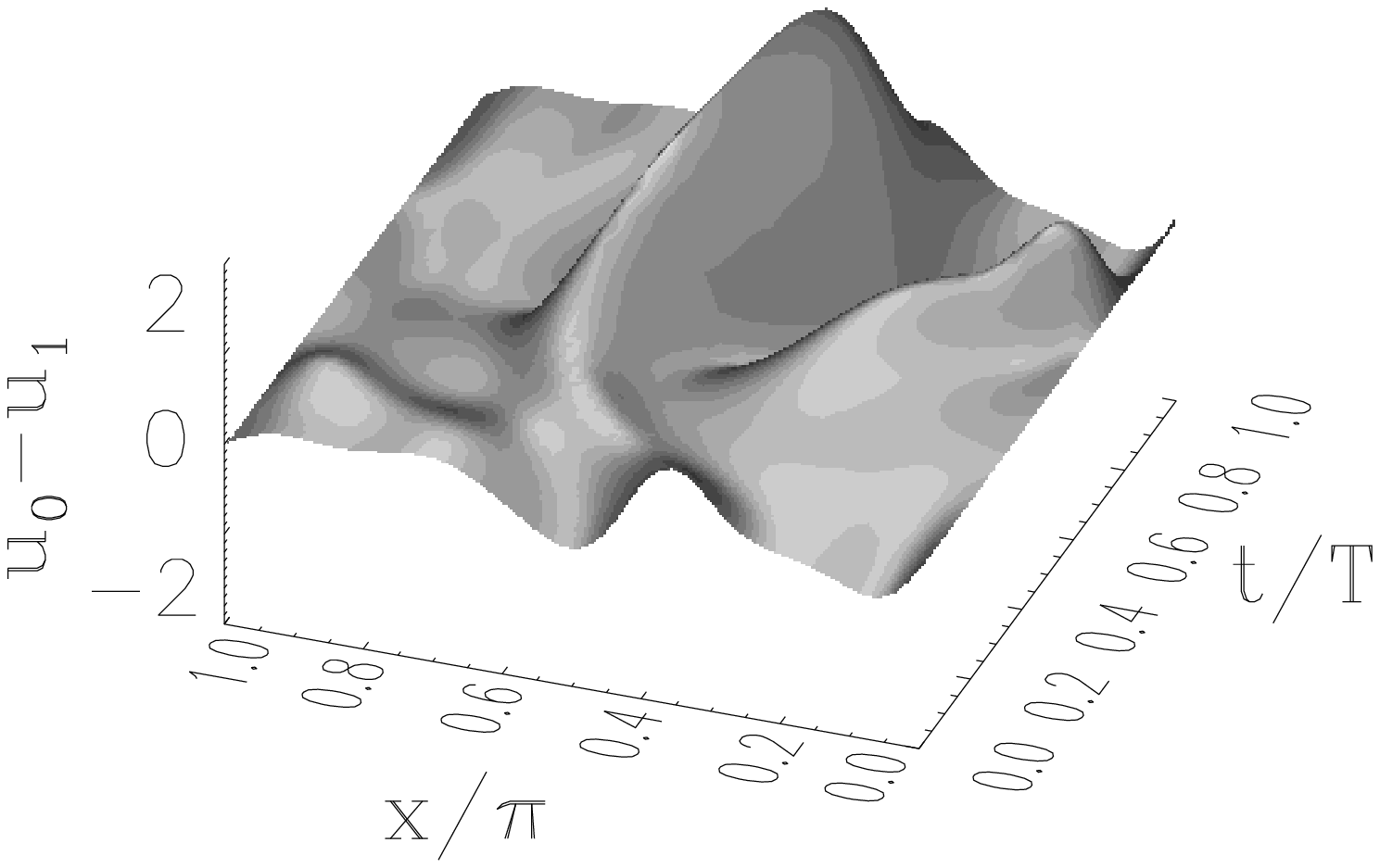}
\hspace{-4ex}
}{}{
(a) Spatio-temporally periodic solution $u_0(x,t)$ of the
Kuramoto-Sivashinsky system,
viscosity parameter $\nu=0.029910$.
(b) The difference between the two shortest period
           spatio-temporally periodic solutions
           $u_0(x,t\period{0})$ and $u_1(x,t\period{1})$.
From \refref{CCP96}.
}{orbit0fig}

\section{Periodic orbit theory} 
\label{PerOrsInAction} 

Now we turn to the central issue; qualitatively, these
solutions demonstrate that the recurrent patterns program
can be implemented, but how
is this information to be used quantitatively? This is 
what the periodic orbit theory is about;
it offers the machinery that assembles
the topological and the quantitative information about individual
solutions
into accurate predictions 
about measurable global averages, such as the Lyapunov exponents
and correlation functions.

Very briefly (for a detailed exposition 
the reader is referred to \refref{QCcourse}), 
the task of any theory that aspires to be a theory of
chaotic, turbulent systems is is to predict the value
of an ``observable"  $\obser$
from the spatial and time averages evaluated along
dynamical trajectories $x(t)$
\[
\left<\obser\right> = \lim_{t\to\infty} {1 \over t} \left<\Obser^t \right>
\,,\qquad 
\Obser^t (x) = \int_0^t \d\tau \, \obser(x(\tau))
\,.
\]
The key idea of the periodic orbit theory
is to extract this average from the leading eigenvalue 
of the {\evOper}
\[
{\Lop}^t (x,y) = \delta(y-x(t))\e^{\beta \Obser^t(x)} 
\]
via the trace formula
\beq
\tr {\Lop}^{t}
     =   \sum_{p}
	\left.
	\raisebox{-4.0ex}[5.5ex][4.5ex]
          {\includegraphics[height=7ex]{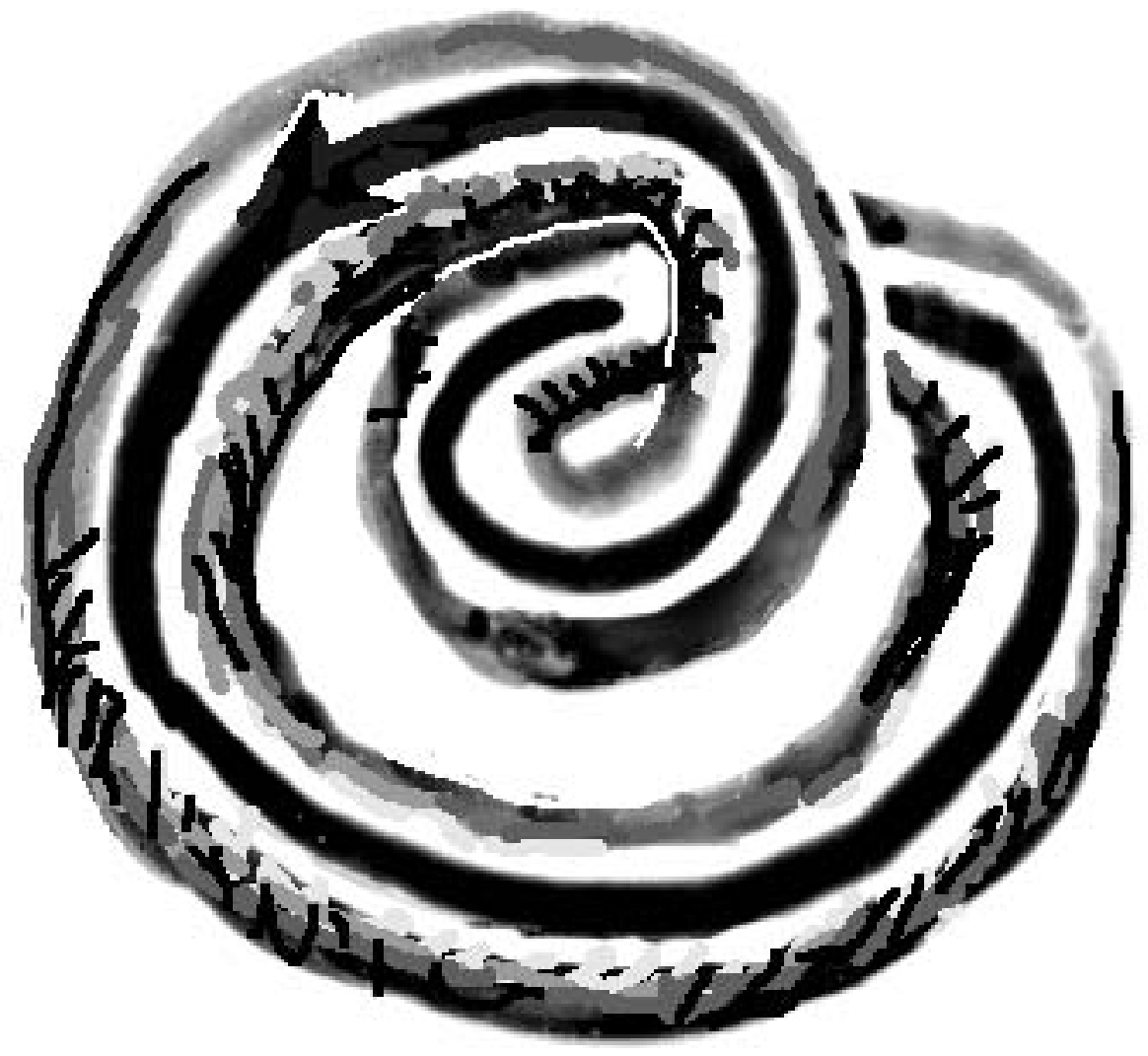}}
	\right._{p}
     =   \sum_{p} \sum_{r=1}^\infty
               { \period{p} \,
                  \prpgtr{t-r \period{p}}
                  \over \oneMinJ{r} }
                \e^{r \beta \Obser_p}
\label{tr-L1}
\eeq
which relates the spectrum of the {\evOper}
to a sum over prime  periodic solutions $p$ of the
dynamical system and their repeats $r$.

What does this formula mean?
Prime cycles partition the dynamical space into neighborhoods,
each cycle enclosed by a
tube whose volume is the product of
its length $\period{p}$ and its thickness
$\left|\det({\bf 1}-{\bf J}_p)\right|^{-1}$.
The trace picks up a periodic orbit contribution only when the
time $t$ equals a prime period or its repeat,
a constraint enforced here by 
$\prpgtr{t- r\period{p}}$. 
${\bf J}_p$ is the linear stability of cycle $p$,
so for long cycles $\oneMinJ{r} \approx$~(product of expanding eigenvalues),
and the contribution of long and very unstable cycles are exponentially
small compared to the short cycles which dominate trace formulas.
The number of contracting directions and the overall dimension
of the dynamical space is immaterial; that is why the theory
can also be applied to PDEs.
All this information is purely geometric, intrinsic to the flow,
coordinate reparametrization invariant,
and the same for any average one might wish to compute.
The information related to a specific observable is carried by the
weight
$\e^{\beta \Obser_p }$, the periodic orbit estimate of the contribution of 
$\e^{\beta \Obser^t(x)}$ from the $p$-cycle neighborhood. 

The intuitive meaning of
a trace formula is that it expresses
the average $\left< \e^{\beta \Obser^t}\right>$ as a 
discretized integral
\vspace{2ex}
\\
\centerline{
	${
	{\mbox{smooth} \atop \mbox{dynamics}}
~~~~~~~~
	\hspace{-4ex}
	\includegraphics[width=0.45\textwidth]{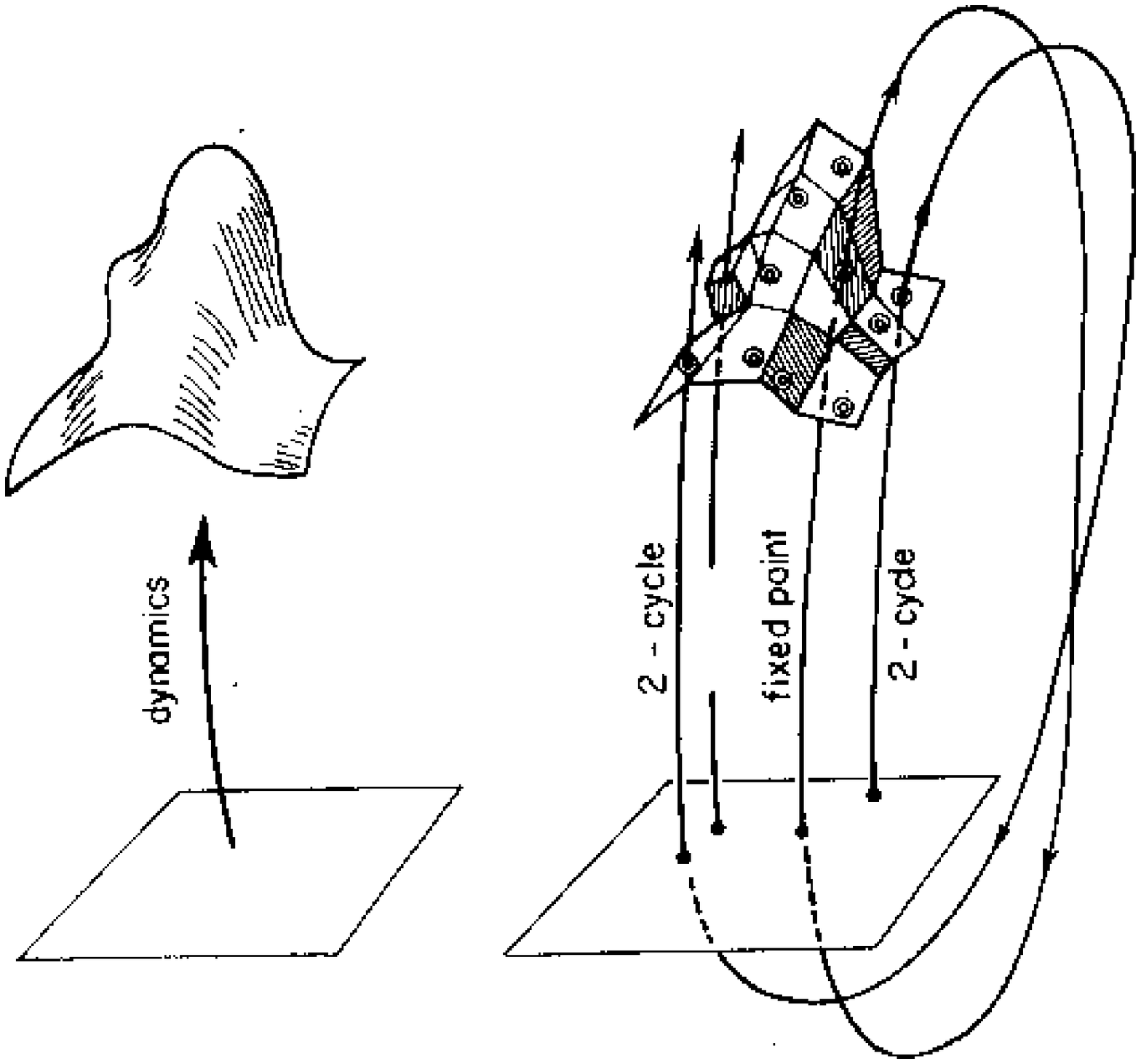}
	\hspace{-4ex}
~~~~~~~~
	{\mbox{linearized}  \atop \mbox{neighborhoods}}
	}$
          }
\vspace{1ex}
\\
over the dynamical space partitioned topologically
into a repertoire of spatio-temporal patterns,
each weighted by the likelihood of pattern's
occurrence in the long time evolution of the system.

Periodic solutions are important because they form the {skeleton}
of the invariant set of the long time dynamics, 
with cycles ordered {hierarchically}; short cycles give good
approximations to the invariant set, longer cycles refinements.
Errors due to neglecting long cycles can be bounded, and for nice hyperbolic
systems they fall off exponentially or even super-exponentially
with the cutoff cycle length\rf{Rugh92}.
Short cycles can be accurately determined and global averages (such as
Lyapunov exponents and escape rates)
can be computed from short
cycles by means of {cycle expansions}.

The Kuramoto-Sivashinsky periodic orbit calculations of
Lyapunov exponents and escape rates\rf{CCP96} demonstrate
that the periodic orbit theory 
predicts observable averages for deterministic but classically
chaotic spatio-temporal systems. 
The main problem today is not how to compute such averages ---
periodic orbit theory as well as direct numerical simulations
can handle that --- but rather that there is no consensus on {\em what}
the sensible experimental observables worth are predicting.

It should be obvious, and it still needs to be said:
the spatio-temporally periodic
solutions are {\em not} to be thought of as eigenmodes,
a good linear
basis for expressing solutions of the equations of motion.
Something like a dilute instant approximation makes no
sense at all for strongly nonlinear systems that we are considering here.
As the equations are nonlinear,
the periodic solutions are in no sense additive,
and their linear superpositions are not solutions. 
\vspace{2ex}
\\
\centerline{
$
	{
A
        \raisebox{-4.0ex}[5.5ex][4.5ex]
                 {
\includegraphics[height=10ex]{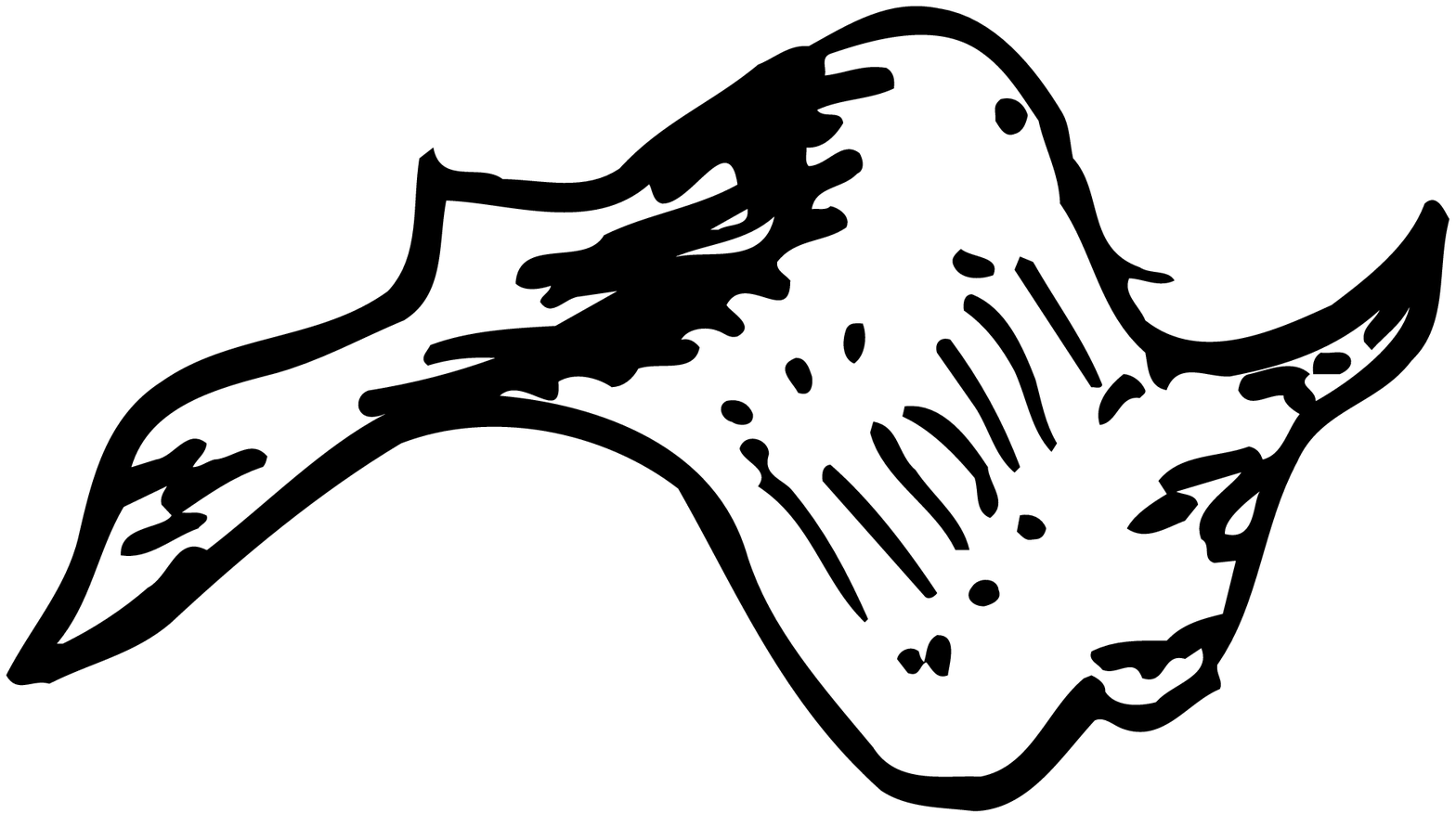}
                 }
~+~ B
        \raisebox{-4.0ex}[5.5ex][4.5ex]
                 {
\includegraphics[height=10ex]{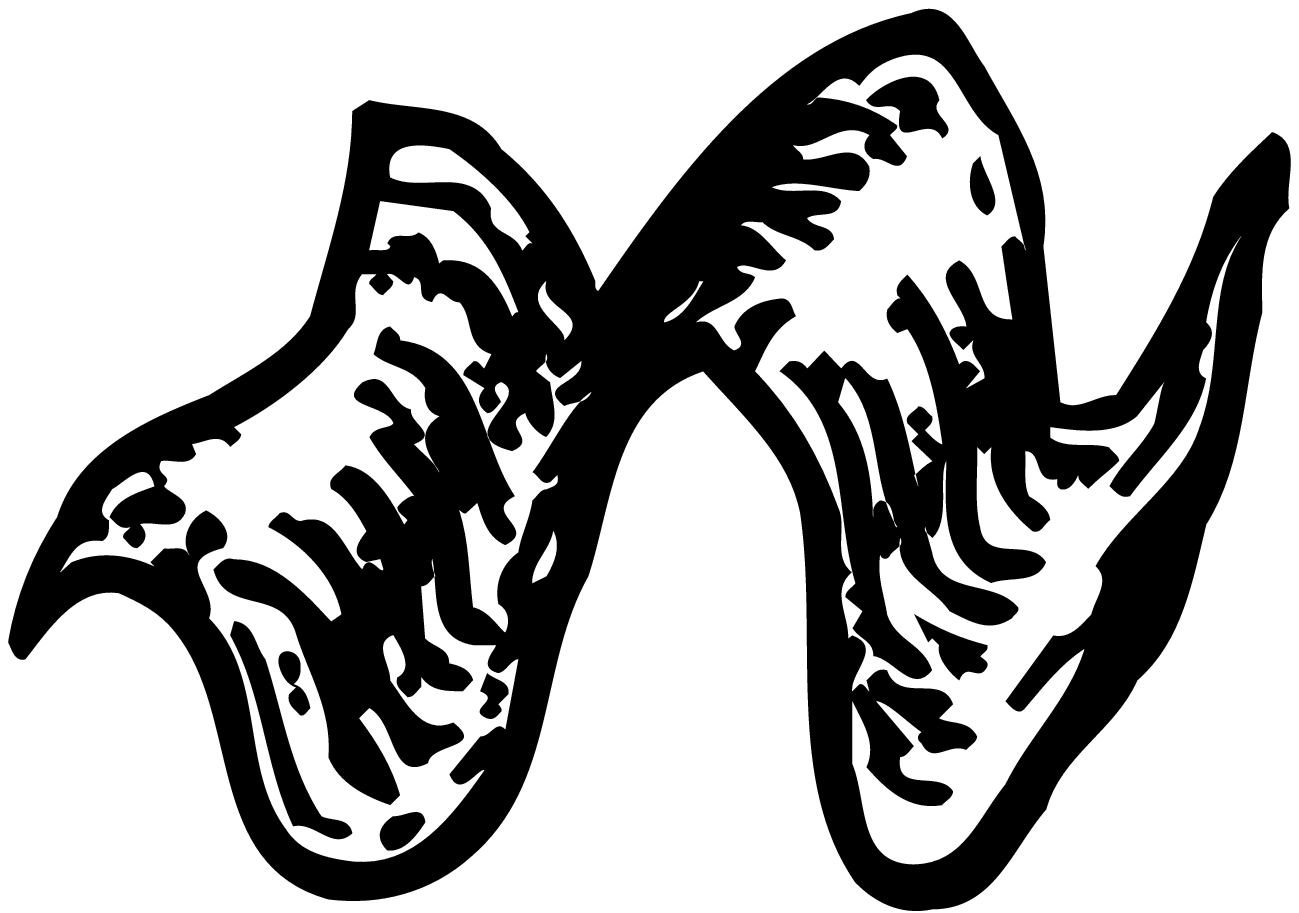}
                 }
+ \,\, \cdots \quad \neq \mbox{~~~{\em u(x,t)}}
        \atop
	\mbox{~~~~~~~~~~~a solution ~~~~~~~~~~~~~~ a solution
              ~~~~~~~~~~~ not a solution}
        }
$
          }
\vspace{2ex}
\\                       
Instead,
it is the trace formulas and {\fd s} of the periodic
orbit theory that prescribe how 
the repertoire of admissible spatio-temporal patterns is
to be systematically explored, 
and how these solutions are to be
put together in order to predict measurable observables. 

Suppose that the above program is successfully carried out for 
classical solutions of some field theory.
What are we to make of this
information if we are interested in the quantum behavior of
the system? In the semiclassical quantization 
the classical solutions are the starting approximation. 

\section{Stochastic evolution}
\label{s-PerCorr}

For the same pragmatic  reasons that we found it profitable
to shy away from facing the 4-dimensional QCD head on in the
above exploratory foray into a strongly nonlinear field
theory, we shall start out  by trying to understand the structure
of perturbative corrections for systems radically simpler than
a full-fledged quantum field theory.
First, instead of perturbative corrections to the quantum
problem, we shall start by exploring the perturbative corrections to
weakly stochastic flows.
Second, instead of continuous time flows, we 
shall start by a study of a discrete time process.

For discrete time dynamics a
Langevin trajectory in presence of additive noise is generated
by iteration
\beq
x_{n+1}=f(x_{n})+\sigma\xi_{n} 
\,,
\ee{mapf(x)-Diag}
where $f(x)$ is a map, 
$\xi_n$ a random variable,
and $\sigma$ parametrizes the noise strength.
In what follows we assume that $\xi_n$ are uncorrelated,
and that the mapping $f(x)$ is
one-dimensional and expanding, but we expect that the form of
the results will remain the same for
higher dimensions, including the field theory example
of the preceding section. 

Tracking an individual noisy trajectory does not make much sense; what
makes sense is the Fokker-Planck formulation, where one
considers instead evolution of an ensemble of trajectories.
An initial density of trajectories $\phi_0(x)$ evolves with time as
\beq
\phi_{n+1}(y) =
\left(
\Lnoise{}
\circ
\phi_{n}\right)(y)
= \int dx \, \Lnoise{}(y,x) \phi_{n}(x)
\ee{DensEvol}
where $\Lnoise{}$ is the {\evOper}
\beq
\Lnoise{}(y,x) 
     =
	\int \delta(y-f(x)-\sigma \xi) P(\xi) d\xi 
  \,=\, \sigma^{-1} P\left[ \sigma^{-1}(y-f(x))\right]
\,,
\label{oper-Diag}
\eeq
and $\xi_n$ a random variable with the normalized
distribution $P(\xi)$, centered on  $\xi = 0$. 

If the noise is weak, the goal of the theory is to compute 
the perturbative corrections to the eigenvalues 
$\eigenvL$ of $\Lnoise{}$
order by order in the noise strength $\sigma$,
\[
\eigenvL(\sigma)
 = \sum_{m=0}^\infty \eigenvL^{(m)}  {\sigma^m \over m!}
\,.
\] 
One way to get at the spectrum of $\Lnoise{}$
is to consider the discrete
Laplace transform of $\Lnoise{}^n$, or the resolvent
\beq
\sum_{n=1}^\infty z^n\tr\Lnoise{}^n
	=
\tr{z\Lnoise{} \over 1 - z\Lnoise{}}
	=
\sum_{\alpha=0}^\infty {z\eigenvL_\alpha \over 1 - z\eigenvL_\alpha}
\ee{discResolv}
which has a pole at every $z= \eigenvL_\alpha^{-1}$.

The effects of weak noise are of interest in their own right, as
any deterministic evolution that occurs in nature is affected by noise.
However, what is most important in the present
context is the fact that the form of perturbative corrections
for the stochastic problem  is the same as for the quantum problem,
and still the actual calculations are
sufficiently simple that one can explore 
many more orders in perturbation theory than would be possible
for a full-fledged field theory, and develop new perturbative
methods.

The first method we try is the standard
Feynman-diagrammatic expansion. For semiclassical quantum mechanics of
a classically chaotic system such calculation was first 
carried out by Gaspard\rf{alonso1}. The stochastic version described
here, implemented by Dettmann\rf{noisy_Fred},
reveals features not so readily apparent in
the quantum calculation.

The Feynman diagram method becomes unwieldy at higher orders.
The second method, introduced by Vattay\rf{diag_Fred},
is based on Rugh's\rf{Rugh92}
explicit matrix representation of the {\evOper}.
If one is interested in evaluating numerically many orders of perturbation
theory and many eigenvalues,
this method is unsurpassed.

The third approach, the smooth conjugations introduced
by Mainieri\rf{conjug_Fred}, is 
perhaps an altogether new idea in field theory.
In this approach
the neighborhood of each saddlepoint is rectified by an appropriate
nonlinear field transformation, with the focus shifted
from the dynamics in the original field variables
to the properties of the conjugacy transformation.  
The expressions obtained are equivalent to
sums of Feynman diagrams, but are more compact.

\section{Feynman diagrammatic expansions}
\label{s-FeynmDiag}

We start our computation of the weak noise corrections to the 
spectrum of $\Lnoise{}$ by calculating the trace of the $n$-th iterate of
the stochastic {\evOper} $\Lnoise{}$.
A convenient choice of noise is Gaussian,
$
P(\xi)=
e^{-\xi^2/2}/{\sqrt{2\pi}}
\,,
$
with the trace given by an $n$-dimensional
integral on $n$ points along a discrete periodic chain
\bea
\tr{\Lnoise{}^n} 
	&=& \int dx_0 \cdots dx_{n-1}
	\Lnoise{}(x_0,x_{n-1}) \cdots \Lnoise{}(x_1,x_0)
	\continue
	&=& \int[dx]\, \exp\left\{-\frac{1}{2\sigma^2}
\sum_{a}\left[x_{a+1}-f(x_a)\right]^2\right\}
	\continue
	&&
x_n  =  x_0 \,,\qquad [dx]=\prod_{a=0}^{n-1}{dx_a \over \sqrt{2\pi\sigma^2}}
\,.
\label{eLnoisMtrx}
\eea
The choice of Gaussian noise is not essential,
as the methods that we develop here apply
equally well to other noise distributions, and more generally to
the space dependent noise distributions $P(x,\xi)$. 
As the neighborhood of any trajectory is nonlinearly
distorted by the flow, the  integrated noise is 
anyway never Gaussian, but colored.

If the classical dynamics is hyperbolic, 
periodic solutions of given finite
period $n$ are isolated.
Furthermore, if the noise broadening $\sigma$ is
sufficiently weak they remain distinct, and the dominant
contributions come from neighborhoods of periodic points, 
the tubes sketched in the trace formula \refeq{tr-L1}.
In the
{\em saddlepoint approximation} the trace \refeq{eLnoisMtrx} is given by
the sum over neighborhoods of periodic points
\beq
\tr{\Lnoise{}^{n}} \longrightarrow
\left. \tr{\Lnoise{}^{n}} \right|_{\mbox{\tiny sc}}
                   =
		\sum_{x_c\inFix{n}} e^{W_c}
  = \sum_p \cl{p} \sum_{r=1}^\infty 
	 \delta_{n, \cl{p} r}  e^{W_{p^r}}
\,.
\ee{SptSum1}
As traces are cyclic, $e^{W_c}$ is the same
for all periodic points in a given cycle, independent of the choice
of the starting point $x_c$,
and the periodic point sum can be rewritten in terms of 
prime cycles $p$ and their repeats.
In the deterministic, $\sigma \to 0$ limit this is the discrete
time version of the classical trace formula \refeq{tr-L1}.
Effects such as noise induced tunnelling are not included in the weak
noise approximation.

We now turn to the evaluation of $W_{p^r}$, the weight of 
the $r$-th repeat of prime cycle $p$.
The contribution of the cycle point $x_a$ neighborhood is best
expressed in an intrinsic coordinate system, by
centering the coordinate system on the cycle points,
\beq
x_a \to x_a + \field_a
\,.
\ee{center-a}
From now on $x_a$ will refer to the position of the $a$-th periodic
point, $\field_a$ to the deviation of the noisy trajectory from
the deterministic one, $f_a(\field_a)$ 
to the map \refeq{mapf(x)-Diag} centered on the $a$-th cycle point,
and $f_a^{(m)}$ to its $m$-th derivative evaluated at the $a$-th cycle point:
\beq
f_a(\field_a) \,= \,f(x_a+\field_a) - x_{a+1}
	\,, \qquad 
f_a'
           =
		f'(x_a)
	   ,\quad
f_a''        =
		f''(x_a)
	   ,\quad\cdots
\,.
\eeq
Rewriting the trace in vector notation, with
$x$ and $f(x)$ $n$-dimensional
column vectors with components $x_a$ and $f(x_a)$
respectively, 
expanding $f$ in Taylor series around each of the periodic points
in the orbit of $x_c$, separating out the quadratic part
and integrating we obtain
\bea
e^{W_c} &=& \int_c[d\field]\, 
     e^{-\left(\InvPrpgtr{}\field_{} - V'(\field) \right)^2/2\sigma^2}
	= \int_c[d\field]\, 
     e^{-{1 \over 2\sigma^2} \,
         \field_{}^T{1 \over {\Prpgtr{}}^T\Prpgtr{}}\field_{}
	\,+\, (\cdots)
       }
	\continue
	&=& |\det \Prpgtr{}| \int_c[d\varphi]\, 
     e^{\sum{1\over k} \tr\left(\Prpgtr{}V''(\field) \right)^k}
     e^{-\varphi^2/2\sigma^2}
\label{eWcMtrx}
\eea
The [$n$$\times$$n$] matrix $\Prpgtr{}$ arises from the quadratic
part of the exponent, while all higher powers of $\field_a$
are collected in $V(\field)$:
\beq
\InvPrpgtr{ab}\field_{b} = -\Df{a}\field_{a}+\field_{a+1}
\,,\qquad
V(\field)= \sum_a \sum_{m=2}^{\infty}f^{(m)}_{a}
\frac{\field_{a}^{m+1}}{(m+1)!}
\,.
\ee{DefPrpg}
The saddlepoint expansion is most conveniently evaluated in terms of
Feynman diagrams, by
drawing $\Prpgtr{}$ as a directed line
$ \Prpgtr{ab} =  \btrack{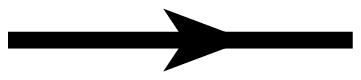}$,
and the derivatives of $V$ as the ``interaction'' vertices
\PC{fix birdtrack height}
\[
f^{''}_a
\,=\,
\btrack{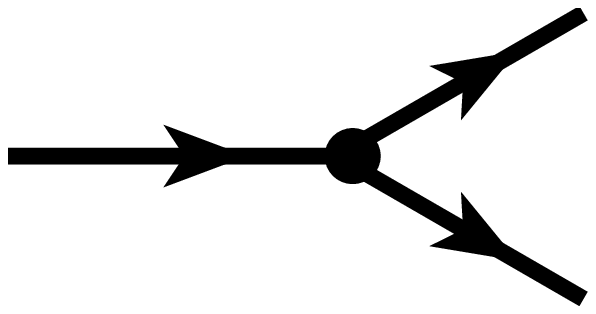}
\,,\quad
f^{'''}_a
\,=\,
\btrack{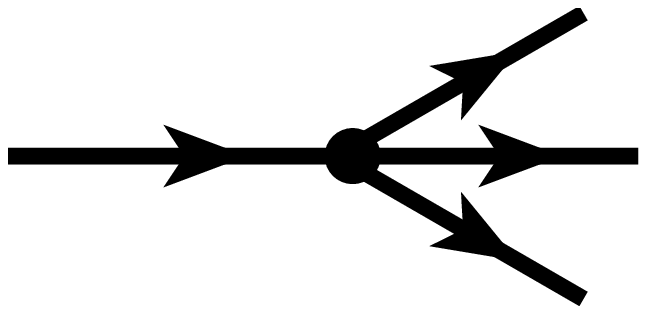}
\,,\quad
\dots
\,.
\]
In the jargon of field theory,
$\Prpgtr{}$ is the ``free propagator''. 
Its determinant
\beq
\left|\det\,\Prpgtr{}\right|={1 \over |\ExpaEig_c-1|}
\,, \qquad
\ExpaEig_c = \prod_{a=0}^{n-1}\Df{a}
\ee{detDel}
is the 1-dimensional version of the classical stability weight
$\left|\det({\bf 1}-{\bf J})\right|^{-1}$
in \refeq{tr-L1},
with $\ExpaEig_c$ the {stability} of the $n$-cycle going through
the periodic point $x_c$. 

Standard methods\rf{FieldThe} 
now yield the perturbation expansion
in terms of the connected ``vacuum bubbles''
\bea
W_c &=& - \ln|\ExpaEig_c-1| + \sum_{k=1}^\infty W_{c,2k}\sigma^{2k}
\label{e:PertExpW}\\
W_{c,2} &=&
	{1 \over 2}\,
	\btrack{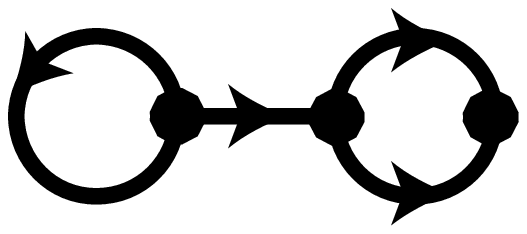}  \,+\,
	{1 \over 2}\,
	\btrack{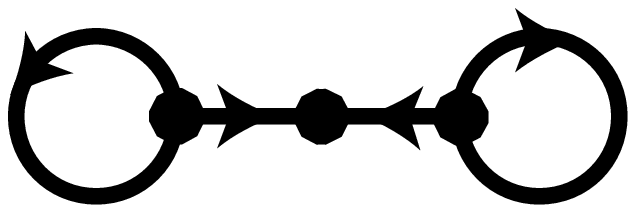} \,+\,
	{1 \over 2}\,
	\btrack{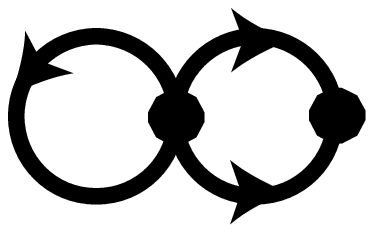} \,+\,
	{1 \over 2}\,
	\btrack{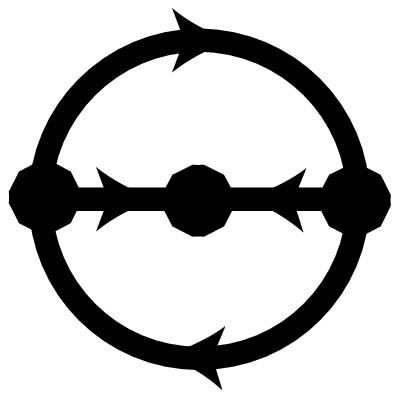}
\label{a:sigSqGraphs} 
	\continue
W_{c,4} &=& \cdots
	\nnu
\,.
\eea
In the usual field-theoretic calculations the $ W_{c,0}$ term
corresponds to an overall volume term that cancels out in the expectation
values. In contrast, as explained in \refsect{PerOrsInAction}, here the
$ e^{W_{c,0}} = |\ExpaEig_c-1|^{-1}$ term 
is  the classical volume of cycle $c$.
 Not only does this weight not 
cancel out in the expectation value formulas, it 
plays the key role both in classical and
semiclassical trace formulas.

In the diagrams sketched above
a propagator line connects $x_a$ at time $a$ with 
$x_{b}$ at later time $b$ by a deterministic trajectory. At time
$b$ noise induces a kick whose strength depends on the local
curvature of the flow. A penalty of a factor $\sigma$ is paid, 
$m-1$ deterministic trajectories originate in 
the neighborhood of $x_{b}$ from vertex $V^{(m)}(x_{b})$,
and the process repeats itself, each vertex carrying a penalty
of $\sigma$, and higher derivatives of the $f_b$.
Summing over all noise kick sequences encoded by
a given diagram and using the periodicity of the trace integral
\refeq{eLnoisMtrx} Dettmann\rf{noisy_Fred} obtains expressions such as
\beq
\frac{r}{2}\frac{\ExpaEig_p^{2r}-1}{\ExpaEig_p^2-1}
 \frac{\ExpaEig_p^r}{(\ExpaEig_p^r-1)^3}
  \sum_{\mod{a}\mod{b}}
    \left(
          \frac{f^{''2}_{\mod{a}}}{f^{'2}_{\mod{a}}}
	 -\frac{f^{'''}_{\mod{a}}}{\Df{\mod{a}}}
    \right)
\prod_{\mod{d}=\mod{b}+1}^{\mod{a}-1}f^{'2}_{\mod{d}}
\,.
\ee{FynmEight}
This particular sum is the $\btrack{FeynmEight}$ 
Feynman diagram $\sigma^2$ correction to $r$-th repeat of prime cycle
$p$.
More algebra leads to similar contributions from
the remaining diagrams. But the overall result is surprising;
the dependence on the repeat number $r$ factorizes, with
each diagram yielding the same prefactor
depending only on $\ExpaEig_p^r$.
This remarkable fact will be explained in \refsect{scfpo}.
The result of the Feynman-diagrammatic
calculations is the {\em stochastic trace formula}
\beq
\left. 
\tr{z\Lnoise{} \over 1 - z\Lnoise{}}
\right|_{\mbox{\tiny sc}} 
	=
\sum_p\sum_{k=0}^{\infty}
	{\cl{p} \, t_{p,k} \over 1-t_{p,k}}
\,,\quad
t_{p,k} 
	=
 \frac{z^\cl{p}}{|\ExpaEig_p|\ExpaEig_p^k}
	  e^{\frac{\sigma^2}{2}    
	     w_{p,k}^{(2)}
 	     + O(\sigma^4)
            }
\,,
\label{NoiseResum}
\eeq
where $t_{p,k}$ is the $k$-th local eigenvalue evaluated on the $p$ cycle.
The deterministic, $\sigma=0$ part of this formula 
is the stochastic equivalent of the Gutzwiller semiclassical
trace formula\rf{gutbook}. The $\sigma^2$ correction $w_{p,k}^{(2)}$ is
the stochastic analogue of Gaspard's $\hbar$ correction\rf{alonso1}.
At the moment the explicit formula
is sufficiently unenlightening that we postpone writing it down to 
\refsect{scfpo}.

While the diagrams are  standard,
the chaotic field theory
calculations are considerably more demanding than is usually the
case in field theory. Here there is 
no translational invariance along the chain, so 
the vertex strength depends on the position, and
the free propagator is not diagonalized by a Fourier transform. 
Furthermore, 
here one is neither ``quantizing'' around a trivial vacuum,
nor a countable infinity of analytically explicit
soliton saddles, but around an
infinity of nontrivial unstable hyperbolic saddles.

Two aspects of the above
perturbative results are {\em a priori} far from obvious:
(a) that the structure of the periodic orbit theory
should survive introduction of noise, and (b) 
a more subtle and surprising result,
repeats of prime cycles can be re-summed and theory reduced to the
\dzeta s and \fd s of the same form as for deterministic
systems.

Pushing the Feynman-diagrammatic approach to higher orders is laborious,
and has not been attempted for this class of problems.
As we shall now see, it is not smart to keep pushing it, either,
as one can compute many more orders of perturbation theory
 by means of a matrix representation for $\Lnoise{}$.

\section{{\EvOper} in a matrix representation}
\label{s-MatrixRep}

An expanding map $f(x)$ takes an initial smooth
distribution $\phi(x)$ defined on a subinterval,
stretches it out and overlays it over a larger interval. 
Repetition of this process smoothes the initial
distribution $\phi(x)$, so it is natural to concentrate on 
smooth distributions $\phi_{n}(x)$, and represent them
by their Taylor series.
By expanding both $\phi_{n}(x)$ and $\phi_{n+1}(y)$ in
\refeq{DensEvol} in Taylor series Rugh\rf{Rugh92} derived a matrix
representation of the {\evOper}
\[
  \int dx \, \Lnoise{}(y,x) \frac{x^{m}}{m!} 
	= \sum_{m'} \frac{y^{m'}}{m'!} \Lmat{}_{m'm}
\,,\qquad m,m'= 0,1,2, \dots
\]
which maps the $x^m$ component of the density of trajectories $\phi_n(x)$ 
in \refeq{DensEvol} to the  $y^{m'}$ component of 
the density $\phi_{n+1}(y)$ one time step later.
The matrix elements follow by differentiating both sides
with $\pde^{m'}/\pde y^{m'}$ and evaluating the integral
\beq
\Lmat{}_{m'm} = 
 \left. \frac{\pde^{m'}}{\pde y^{m'}}
  \int dx \, \Lnoise{}(y,x)
\frac{x^{m}}{m!} \right|_{y=0} 
\,.
\ee{Lmat-Diag}



In \refeq{oper-Diag} we have written the {\evOper} $\Lnoise{}$ in terms of
the Dirac delta function
in order to emphasize that
in the weak noise limit the stochastic trajectories are concentrated 
along the classical trajectory $y = f(x)$.
Hence it is natural to
expand the kernel in a Taylor series~\rf{Watanabe87}
in $\sigma$
\beq
\Lnoise{}(y,x) =
	\delta(y-f(x)) 
	+\,
\sum_{n=2}^{\infty}\frac{(-\sigma)^n}{n!}\delta^{(n)}(y-f(x))
\int \xi^n P(\xi)d\xi  
\,,
\ee{opexp}
where
$
	\delta^{(n)}(y) = {\pde^n \over \pde y^n} \delta(y)
	\,.
$
This yields a representation of the
{\evOper} centered along the classical trajectory, 
dominated by  the  deterministic
{\FPoper} $\delta(y-f(x))$, with corrections given by derivatives
of delta functions weighted by  moments of the noise
distribution $P_n=\int P(\xi)\xi^nd\xi$.
We again center the coordinate system on the cycle points
as in \refeq{center-a},
and also introduce a notation for the operator \refeq{oper-Diag}
centered on the $x_a \to x_{a+1}$ segment of the
classical trajectory
\[
\Lnoise{a}(\field_{a+1},\field_a)
           \,=\,
   \Lnoise{}(x_{a+1}+\field_{a+1},x_a+\field_a)
\,.
\]
The weak noise expansion \refeq{opexp} for the $a$-th segment operator
is given by
\beq
\Lnoise{a}(\field_{a+1},\field_a)=
	\delta(\field_{a+1}- f_a(\field_a))
	+\sum_{n=2}^{\infty}\frac{(- \sigma)^{n}}{n!}
	P_n \delta^{(n)}(\field_{a+1}- f_a(\field_a))
\,.
\label{LnoiseExp}
\eeq
As the {\evOper} has a simple $\delta$-function form,
the local matrix representation of $\Lnoise{a}$ centered on
the $x_a \to x_{a+1}$ segment of the deterministic trajectory
can be evaluated recursively in terms of derivatives of the map $f$:
\bea
 \left(\Lmat{a}\right)_{m'm} 
	&=& 
 \sum_{n}^{\infty}
P_n\frac{(-\sigma)^n}{n!}(\Bmat{a})_{m'+n,m}
 \,,\qquad n= \mbox{max}(m-m',0)
	\continue
(\Bmat{a})_{m' m}
	&=& 
    \int d\field \, 
     \delta^{(m')}(\field_{a+1} -f_a(\field)) \frac{\field^{m}}{m!}
\label{Bmatrix}\\
	&=& 
	\left.
	{1 \over |f_a'|}
	\left( {d~ \over d\field}  {1 \over f_a'(\field)}\right)^{m'}
        \frac{\field^{m}}{m!}
        \right|_{\field=0}
\,.
\nnu
\eea
The matrix elements vanish for $m'<m$, so $\Bmat{}$ is 
a lower triangular matrix. 
The diagonal and the successive off-diagonal matrix elements 
are easily evaluated iteratively by computer algebra
\[
(\Bmat{a})_{m m} = \frac{1}{|f_a'|(f_a')^{m}}
	\,,\quad
(\Bmat{a})_{m+1,m} = - \frac{(m+2)! f_a''}{ 2 m! |f_a'|(f_a')^{m+2}}
\label{Bexplicit}
\,,~~\cdots
\,.
\]
For chaotic systems the map is expanding, $|f_a'|>1$. Hence
the diagonal terms drop off exponentially, as $1/|f_a'|^{m+1}$,
the terms below the diagonal fall off even faster, and 
truncating $\Lmat{a}$ to a finite matrix 
introduces only exponentially small errors.

The trace formula \refeq{discResolv}
takes now a matrix form
\beq
\left.
\tr{z\Lnoise{} \over 1 - z\Lnoise{}}
\right|_{\mbox{\tiny sc}}
        =
\sum_p \cl{p} \tr {z^\cl{p} \Lmat{p} \over 1-z^\cl{p} \Lmat{p} }
\,,
\label{tracenp}
\eeq
where
$ \Lmat{p}  = { \Lmat{\cl{p}}\Lmat{2}\cdots \Lmat{1}}$
is the contribution of the $p$ cycle.
The subscript {\tiny sc} is a reminder that this is a saddlepoint
or semiclassical approximation,
valid as an asymptotic series in the limit of weak noise.
Vattay\rf{VatBS}
interprets the local matrix representation of the \evOper\ as follows.
The matrix identity log~det~=~tr~log together with 
the trace formula \refeq{tracenp} yields
\bea
\left. 
 \det(1 - z \Lop)
\right|_{\mbox{\tiny sc}} 
	=
 \prod_{p} \det(1 - z^{\cl{p}} \Lmat{p}) 
\,,
\label{fredholmprod}
\eea
so in 
the saddlepoint approximation the spectrum of the {\em global}
{\evOper} $\Lnoise{}$ is 
pieced together from the {\em local} spectra computed cycle-by-cycle
on neighborhoods of individual prime cycles with periodic boundary
conditions. The meaning of the $k$-th term in 
the trace formula \refeq{NoiseResum} is now clear; it is the
$k$-th eigenvalue of the local {\evOper} restricted to the
$p$-th cycle neighborhood.

Using this matrix representation Palla and S\o ndergaard\rf{diag_Fred}
were able to compute corrections to order $\sigma^{12}$,
a feat simply impossible along the Feynman-diagrammatic
line of attack.
In retrospect, the matrix representation method for solving the
stochastic evolution is eminently sensible --- after all, that is 
the way one solves a close relative to stochastic PDEs,
the Schr\"odinger equation. What is
new is that the problem is being solved locally, 
periodic orbit by periodic orbit, by translation to 
coordinates intrinsic to the periodic orbit. It
is this natural local basis that makes the 
matrix representation so simple.

Mainieri\rf{conjug_Fred}
takes this observation one step further; as the dynamics is 
nonlinear, why not search for a nonlinear coordinate transformation that
makes the intrinsic coordinates as simple as possible?

\section{Smooth conjugacies}
\label{scfpo}

This step 
injects into field theory a method
standard in the construction of normal forms for bifurcations\rf{KatHass}.
The idea is to perform a smooth nonlinear coordinate transformation
$x = h(y)$,
$ f(x) = h(g(h^{-1}(x)))$
that flattens out the vicinity
of a fixed point and makes the map {\em linear} in an open neighborhood,
$ f(x) \to g(y) = {\bf J} \cdot y$.
\vspace{2ex}
\\
\centerline{
 ${
	\raisebox{-4.0ex}[5.5ex][4.5ex]
		 {
\includegraphics[height=10ex]{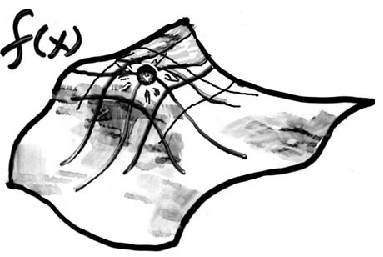}
		 }
        \atop
        \mbox{an arbitrary coordinatization}
        }$
~~~
$\Longrightarrow$
~~~
 ${
	\raisebox{-4.0ex}[5.5ex][4.5ex]
		 {
	\includegraphics[height=10ex]{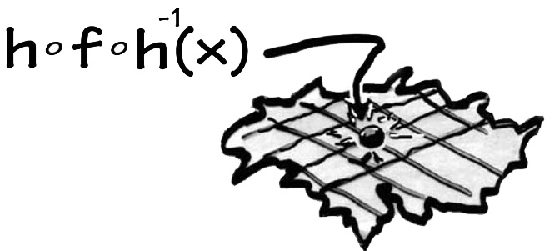}
		 }
        \atop
        \mbox{intrinsic, flat coordinates}
        }$
          }
\vspace{2ex}
\\
The key idea of flattening the neighborhood of
a saddlepoint can be traced back to 
Poincar\'e's celestial mechanics,
and is perhaps not something that a field theorist would instinctively
hark to as a method of computing perturbative corrections.
This local rectification of a map can be implemented
only for isolated non-degenerate fixed points (otherwise
higher terms are required by the normal form expansion around the
point), and only in finite neighborhoods, as the 
conjugating functions in general have finite radia of convergence.

We proceed in two steps. First, substitution of
the weak noise perturbative
expansion of the {\evOper} \refeq{LnoiseExp} into the trace
centered on cycle $c$ generates products of derivatives of
$\delta$-functions:
\[
\left. \tr{\Lnoise{}^{n}} \right|_{c}
= \cdots + \int[d\field]\, 
	\left\{\cdots 
	\delta^{(m')}(\field''- f_a(\field'))
	\,
	\delta^{(m)}(\field'- f_{a-1}(\field))
	\cdots \right\}
	+ \cdots 
\,.
\]
The integrals are evaluated as in \refeq{Bmatrix}, yielding
recursive derivative formulas such as
\beq
    \int dx \, 
     \delta^{(m)}(y)
	=
	\left.
	{1 \over |y'(x)|}
	\left(- {d~ \over dx}  {1 \over y'(x)}
	\right)^{m}
        \right|_{y=0}
     \,,\qquad
	y=f(x)-x
\,.
\ee{recurDer}
or $n$-point integrals, with derivatives distributed
over $n$ different $\delta$-functions.

Next we linearize the neighborhood of the $a$-th cycle point.
For a 1-dimensional map $f(x)$ with a fixed point $f(0)=0$
of stability $\ExpaEig = f'(0)$, $|\ExpaEig|\neq 1$ we search
for a smooth conjugation $h(x)$
such that: 
\beq
	f(x) = h(\ExpaEig h^{-1}(x))
\,,\quad
h(0)=0
\,,\quad
h'(0)=1
\,.
\label{EqConj}
\eeq
In higher dimensions $\ExpaEig$ is replaced by the Jacobian matrix
${\bf J}$.
For a periodic orbit each point around the cycle has a
differently distorted neighborhood,
with differing second and higher derivatives,
so the conjugation function $h_a$ has to be computed
point by point, 
\[
f_a(\phi)=h_{a+1}(f'_a h_a^{-1}(\phi))
\,.
\]
An explicit expression for $h_a$ in terms of $f$ is 
obtained by iterating around the whole cycle, and using the chain
rule \refeq{detDel} for the cycle stability $\ExpaEig_p$
\beq
f_a^{\cl{p}}(\phi)=h_a(\ExpaEig_p h_a^{-1}(\phi))
\,,
\ee{hCycle}
so each $h_a$ is given by some combination
of $f_a$ derivatives along the cycle.
Expand $f(x)$ and $h(x)$
\[
	f(x) 
	=
	\ExpaEig x + x^2 f_2 + x^3f_3 + \dots
	\,,\quad
	h(y) 
	=
	y + y^2 h_2 + y^3 h_3 + \dots
	\;\,,
\]
and equate recursively coefficients in the
functional equation $ h(\ExpaEig y)=f(h(y))$ expansion
\beq
	h(\ExpaEig u) - \ExpaEig h(u) = 
		\sum_{n=2}^\infty f_m \left(h(u)\right)^m
	\;\,.
\label{EqConjExp}
\eeq
This yields the expansion 
for the conjugation function $h$ 
in terms of the mapping $f$
\beq
h_2 = \frac{f_2}{ \ExpaEig (\ExpaEig - 1) }
	\,,\qquad
h_3 =
\frac{2 f_2^2 + \ExpaEig  ( \ExpaEig -1  )   f_3}
   { \ExpaEig^2 
        ( \ExpaEig -1 ) 
        ( \ExpaEig^2 -1 ) 
   }
\,, \qquad\cdots
	\;\,.
\ee{h(2)}
The periodic orbit
conjugating functions $h_a$ are obtained in the same way
from \refeq{hCycle},
with proviso that
the cycle stability is not marginal, $|\ExpaEig_p|\neq 1$.

What is gained by replacing the perturbation expansion 
in terms of $f^{(m)}$ by still
messier perturbation expansion for the
conjugacy function $h$?
Once the
neighborhood of a fixed point is linearized, the conjugation formula
for the repeats of the map 
\[
	f^r(x)=h(\ExpaEig^r h^{-1}(x))
\]
can be used to compute
derivatives of a function composed with itself $r$ times.
The expansion for arbitrary number of repeats depends
on the conjugacy function $h(x)$
computed for a {\em single} repeat, and all the dependence on the
repeat number is carried by polynomials in $\ExpaEig^r$,
a result that emerged as a surprise in the Feynman diagrammatic
approach of \refsect{s-FeynmDiag}.
The integrals such as \refeq{recurDer} evaluated on the $r$-th repeat of
prime cycle $p$
\beq
y(x) =  f^{\cl{p}r}(x) - x
\ee{g(x)}
have a simple dependence
on the conjugating function $h$
{\small
\bea
\frac{1}{3!}\frac{\partial^2}{\partial y^2}
\frac{1}{y'(0)}
	&=&
	{\frac{{{\ExpaEig }^r}\left( 1 + {{\ExpaEig }^r} \right) 
	}{{{\left( \ExpaEig^r -1 \right)
          }^3}}}
      \left(2h_2^2 -  h_3 \right)
\label{der(5)} \\
	\continue
\frac{1}{4!}\frac{\partial^3}{\partial y^3}
	\frac{1}{y'(0)}
	&=&
	-5\ExpaEig^r{ {( \ExpaEig^r + 1 )^2}
	    \over
            (\ExpaEig^r - 1)^4 }h_2^3 
        + \ExpaEig^r { 5\ExpaEig^{2r} + 8\ExpaEig^r + 5
	    	      \over
            	      (\ExpaEig^r - 1)^4 } h_2h_3 
         - \ExpaEig^r { \ExpaEig^{2r} + \ExpaEig^r + 1
                      \over
                      (\ExpaEig^r - 1)^4 } h_4
	\continue
\cdots	&=&\cdots
\nnu
\eea
}
The evaluation of $n$-point integrals is more subtle\rf{conjug_Fred}.
The final result of all these calculations is that 
expressions of form \refeq{der(5)} depend
on the conjugation function determined from the
iterated map,
with the saddlepoint approximation to
the {\fd} given by
\[
\left. \det(1-z{\Lop}_\sigma)\right|_{\mbox{\tiny sc}}
	=
	 \prod_p\prod_{k=0}^{\infty}(1-t_{p,k})
\]
in terms of local $p$-cycle eigenvalues
\bea
t_{p,k}&=&\frac{z^{\cl{p}}}{|\ExpaEig_p|\ExpaEig_p^k}
\e^{\frac{\sigma^2}{2}P_2 w^{(2)}_{p,k}
    +\frac{\sigma^3}{3!}P_3 w^{(3)}_{p,k}
    +\frac{\sigma^4}{4!}P_4 w^{(4)}_{p,k}+O(\sigma^6)}
	\continue
w^{(2)}_{p,k}&=&(k+1)^2\sum_a (2h_{a,2}^{2} - h_{a,3})
	\,,\qquad
w^{(3)}_{p,k} = \cdots
	\,,\cdots \,.
\nnu
\eea
accurate up to order $\sigma^4$.
$w^{(3)}$, $w^{(4)}$ are also computed by Dettmann,
but we desist from citing them here; the reader is
referred to
\refref{conjug_Fred}. 
What is remarkable about these results is their simplicity when
expressed in terms of the conjugation function $h$, as opposed
to the Feynman diagram sums, in which each diagram contributes
a sum like the one in \refeq{FynmEight}, or worse. Furthermore, both
the conjugation and the matrix approaches are easily automatized,
as they require only recursive evaluation of derivatives, as
opposed to the handcrafted Feynman diagrammar.

Simple minded as they might seem, discrete stochastic processes are
a great laboratory for testing ideas that would otherwise be hard to
test. Dettmann, Palla and S\o ndergaard have 
used a 1-dimensional repeller of bounded nonlinearity
and complete binary symbolic dynamics to check numerically
the above results, and computed the leading eigenvalue of
$\Lnoise{}$ by no less than five different methods.
As anticipated by Rugh\rf{Rugh92},
the {\evOper} eigenvalues
converge super-exponentially with the cycle length; addition
of cycles of period $(n$+1$)$ to the set of all cycles
up to length $n$ {\em doubles} the number of significant 
digits in the perturbative prediction.
However, as the series is asymptotic, for realistic values 
of the noise strength summations beyond all orders are
needed\rf{asym_Fred}.

\section{Summary}

The periodic orbit theory approach to turbulence is to visualize 
turbulence as a sequence of near recurrences in a repertoire of unstable 
spatio-temporal patterns. 
The investigations  of the Kuramoto-Sivashinsky system
discussed above are first
steps in the direction of implementing this program. 
So far, existence of a 
hierarchy of spatio-temporally periodic solutions 
of spatially extended nonlinear system has been demonstrated,
and the periodic orbit theory has been tested in
evaluation of global averages for such system.
The parameter ranges tested so far probe
the weakest nontrivial ``turbulence'', and it is an open 
question to what extent the approach remains implementable 
as the system goes more turbulent. 

The most important lesson of this investigation is that 
the unstable spatio-temporally periodic
solutions do explore systematically the
repertoire of admissible spatio-temporal patterns, with the trace
and {\fd s} formulas and their cycle expansions being the proper tools for
extraction of quantitative predictions from the periodic orbits data.

We formulat next a semiclassical perturbation theory for
stochastic trace formulas 
with support on infinitely many chaotic saddles.
The central object of the periodic orbit theory, the trace of the {\evOper},
is a discrete path integral, similar to those found in field theory
and statistical mechanics.  The weak noise perturbation theory,
likewise, resembles perturbative field theory, and can be cast
into the standard field-theoretic language of Feynman diagrams.
However, we found out that both the matrix and the nonlinear
conjugacy perturbative methods are superior to the standard approach.
In contrast to previous perturbative expansions around
vacua and instanton solutions, the location
and local properties of each saddlepoint must
be found numerically.  

The key idea in the new formulation of perturbation
theory is this: Instead of separating the action into 
quadratic and ``interaction'' parts, one first performs a nonlinear 
field transformation which turns the saddle point into an exact
quadratic form.  The price one pays for this is the Jacobian of
the nonlinear field transformation ---
but it turns out
that the perturbation expansion of this Jacobian 
{in terms of the conjugating function} is order-by-order
more
compact than the Feynman-diagrammatic expansion.

\section*{Acknowledgements}
I am indebted to my collaborators
C.P.~Dettmann, G.~Vattay,
F.~Christiansen,
V.~Putkaradze,
G.~Palla, N.~S\o ndergaard, 
R.~Mainieri
and 
H.H.~Rugh
for co-suffering through all the details omitted in
this overview.
I am grateful to 
E.A. Spiegel, L.~Tuckerman and M.J.~Feigenbaum
for patient instruction. I am not grateful to those
directors and gentlemen of committees who
do not find theoretical physics a
vibrant subject.





\end{document}